\documentclass[12pt]{article}
\usepackage{epsfig}
\usepackage{graphicx}

\usepackage{amssymb,amsmath}

\def\la{\langle }
\def\ra{ \rangle }
\def\la{\langle }

\newcommand{\beq}{\begin{equation}}
\newcommand{\eeq}{\end{equation}}
\newcommand{\bea}{\begin{eqnarray}}
\newcommand{\eea}{\end{eqnarray}}

\begin{document}
\begin{titlepage}
\begin{flushright}
\end{flushright}
\vskip3cm
\begin{center}
{\LARGE
Climbing Down from the Top:
\vskip0.5cm
Single Name Dynamics in Credit Top Down Models} 
\vskip1.0cm
{\Large Igor Halperin and Pascal Tomecek} 
\vskip0.5cm
Quantitative Research, JP Morgan 
\vskip1.0cm
September 1, 2008
\vskip1.0cm
\vskip1.0cm
{\Large Abstract:\\}
\end{center}
\parbox[t]{\textwidth}{

In the top-down approach to multi-name credit modeling, calculation of 
singe name sensitivities appears possible, at least in principle, within the 
so-called random thinning (RT) procedure which dissects the portfolio risk into 
individual contributions. We make an attempt to construct a practical RT 
framework that enables efficient calculation of 
single name sensitivities in a top-down framework, and can be extended to valuation
and risk management of bespoke tranches. 
Furthermore, we propose a dynamic extension of the RT method that
enables modeling of both idiosyncratic and default-contingent individual spread
dynamics within a Monte Carlo setting in a way that preserves the portfolio 
``top''-level dynamics. This results in a model that is not only 
calibrated to tranche and single name spreads, but can also be tuned to 
approximately match given levels of spread volatilities and correlations of names
in the portfolio.    
}

\vspace{2.0cm}
\newcounter{helpfootnote}
\setcounter{helpfootnote}{\thefootnote} 
\renewcommand{\thefootnote}{\fnsymbol{footnote}}
\setcounter{footnote}{0}
\footnotetext{ 
We would like to thank Andrew Abrahams, Anil Bangia, Tom Bielecki, Rama Cont, Kay Giesecke, Andrei Lopatin,
Philipp Sch{\"o}nbucher, Jakob Sidenius, Tolga Uzuner, Yury Volvovskiy and Michael Walker for 
very helpful discussions. 
All remaining errors are our own.
} 
\renewcommand{\thefootnote}{\arabic{footnote}}
\setcounter{footnote}{\thehelpfootnote} 

                                                \end{titlepage}

\section{Introduction}

Modeling of a dynamic credit portfolio can proceed according to two different 
paradigms. In a {\it bottom-up} approach, we start with single name dynamics which are 
constructed to fit individual credit spreads. At the second stage, one attempts to 
calibrate the dependence structure (introduced via common factors or a copula) to 
portfolio pricing data such as tranche prices. However,
calibration to tranches across multiple strikes and maturities 
can be quite challenging in this framework. Furthermore, the
bottom-up approach makes it difficult (though not impossible, see 
Collin-Dufresne et al (2003) \cite{Dufresne} 
and Sch{\"o}nbucher (2003) \cite{Schon1}) to
introduce default clustering (contagion) in a tractable way.

In many applications, the portfolio loss (or spread-loss) 
process is all we need for pricing and 
risk management. In particular, this is the case when one
wants to price a non-standard index tranche off standard ones, and 
risk-manage it using the same standard tranches plus the credit index.
Another example is an exotic 
portfolio
derivative (such as e.g. a tranche option) that is hedged using tranches written on 
the same portfolio. In recognition of this fact, 
the alternative, {\it top-down} approach, suggested by 
Giesecke and Goldberg (GG) (2005) \cite{Giesecke1}, suggests that a 
portfolio-level (or economy-level) loss process, rather 
than loss processes for individual names, should be the modeling primitive. 
Such a process is generally much easier to calibrate to tranche prices than 
an aggregate
portfolio loss process obtained with the bottom-up approach. 
Moreover, credit contagion is built-in in this approach by construction 
\cite{Giesecke1, Giesecke2}. Once a calibrated 
portfolio loss process is obtained, it can be used to price vanilla and 
exotic derivatives referencing the same portfolio. (For a further comparison between the 
bottom-up vs top-down paradigms and a brief review of published models,
see e.g. Bielecki, Cr{\'e}pey and Jeanblanc \cite{BCJ} and references therein.)

The two alternative approaches to modeling credit portfolios, i.e. the 
bottom-up and top-down paradigms, are somewhat complimentary in the sense that
typically, for practical applications we chose one of them based on its appropriateness as 
well as practicality of its use. If we are interested in pricing (and/or risk management
using only a credit index and its tranches), then a 
transition to a coarse-grained description of the top-down approach can be justified,
as eventually our payoffs are only functions of portfolio-level losses, not of losses on
individual names.    
Note, however, that the knowledge of the portfolio loss process is 
not sufficient in more complicated (but common) 
situations, e.g. when a risk manager chooses to use single names for
hedging. Hedging on a single name basis is often done for names in the portfolio
whose spreads significantly widen. Obviously, calculation of single name sensitivities
is beyond the reach of a model whose only output is a portfolio-level loss process.



In addition to the hedging issue, there exists another reason why single name 
calibration can be 
of substantial practical interest, which is the so-called 
``bespoke problem''. The problem is 
that most exotic portfolio credit derivatives (forward-starting tranches, tranche 
options, etc) 
reference customized 
(bespoke) portfolios, rather than standardized index portfolios. If we work 
with a top-down 
model, we have to 
calibrate it to tranches referencing this bespoke portfolio. However, such 
portfolios usually 
do not have enough 
liquidity, which forces practitioners to mark prices of tranches on bespoke 
portfolios off 
tranche quotes for 
index portfolios, using some sort of a mapping between the two portfolios. 
Currently, the 
market standard is to
use the base correlation methodology along with an adjustment to the difference 
in expected 
portfolio losses between
the two portfolios. However, this procedure is quite {\it ad hoc} and is neither 
consistent 
nor satisfactory. In 
particular, it often violates no-arbitrage conditions, thus posing 
substantial practical 
difficulties. 

The bottom-up approach suggests, in principle, a natural method to deal with the bespoke 
problem: once a model is calibrated to market data at both 
the single name and tranche levels, arbitrary bespoke tranches 
can be priced with the same model by first choosing parameters obtained by calibration to index 
tranche prices for names that are common for the bespoke and index portfolios, and then relying on 
interpolation/extrapolation for names that are not.
However, this argument can also be applied to a top-down model once we can calibrate it to single names in addition to tranches. In this case, the bespoke problem would be resolved in the 
top-down approach in exactly the same way as in the bottom-up approach.  

With the top-down approach, information about single names can be recovered,
at least in principle, using the so-called {\it random thinning} (RT) procedure
proposed by Giesecke and Goldberg (2005) \cite{Giesecke1}. The idea of this approach
is to allocate fractions of portfolio default intensity calibrated within the 
``top'' part of the model to individual names in such a way that individual CDS
spreads of names in the portfolio are matched.
Once this is done, the model 
becomes calibrated to both tranche and single name data. 
Various versions of the RT discussed in 
\cite{Giesecke1,Giesecke2,Giesecke3} use thinning of the portfolio intensity by
deterministic or piecewise-deterministic (ones that jump upon defaults in 
the portfolio) processes. (This involves a number of subtle technical points,
see BCJ \cite{BCJ} and also below in Sect.2.3.)

In this paper, we take a probabilistic view of the problem of construction of 
single name dynamics within the top-down framework. Note that the ``top'' part
of a top-down model can be viewed as producing 
``incomplete information'' scenarios that 
yield a forecast of the timing of sequential defaults in the portfolio, but lose
information on the defaulters' identities. In a sense, our problem thus
becomes a problem of {\it inference}, where we probabilistically assign 
identities to all future defaulters. 
As will be clear in what follows, our approach bears resemblance to
the well-known statistical problem of inference of a two-dimensional 
(2D) probability distribution when only its 1D marginals (corresponding, in our
case, to 
portfolio-level and individual names default forecasts, respectively) are known. 

The contribution of this paper is two-fold. First, 
we present a practical RT framework that enables efficient calculation of 
single name sensitivities in a top-down framework, and can be extended to valuation
and risk management of bespoke tranches. Second, we suggest a way of 
extending this framework to model {\it dynamics} of single names 
in a way that preserves both the portfolio-level calibration and dynamics. This is 
achieved by formulating and solving the problem of single name dynamics as 
a filtering problem\footnote{ 
The whole construction developed to this end can be viewed as an 
``inverse dynamic copula'': while 
a dynamic copula constructs a multivariate dynamic process in a way consistent with dynamic 
marginals, it is exactly the other way around in our construction.}.
We construct an "information process" to explicitly model the market filtration, and further show how parameters of this process could be chosen to approximately (in the portfolio-averaged sense) 
calibrate
to a given set of single name spread volatilities and correlations. We note in this
relation that most, if not all, of credit portfolio models usually concentrate on 
matching individual CDS spreads of names in a portfolio as well as tranche spreads,
but not finer characteristics such as spread volatility. However, it can be 
expected that better control of spread dynamics might be important for the pricing
and risk management of exotic portfolio credit derivatives. The particular 
framework employed in our construction does allow one to have some
control of single name spread dynamics, which thus might be seen as an attractive 
feature of the top-down approach. 
 
The paper is organized as follows. In Sect.\ref{SectionTD} we introduce the so-called top-down
(TD) matrices which serve as the basic modeling primitives of our 
approach. Sect.\ref{SectionThinning} then
introduces the iterative scaling algorithm for calculation TD-matrices 
calibrated to single names. In Sect.\ref{SectionSens} we calculate single name sensitivities with 
our approach. Sect.\ref{SectionDynamics} describes the dynamics extension of our RT framework.
A very short Sect.\ref{SectionNumerical} summarizes the numerical Monte Carlo algorithm for simulation
single name dynamics. In Sect.\ref{SectionMarking} we outline a method of approximate calibration 
of parameters driving the single name dynamics. The final Sect.\ref{SectionSummary} presents some 
discussion and conclusions.
 
\section{Top-down default time matrices}\label{SectionTD}

The following notation will be used throughout this paper:
\begin{itemize}
\item $ \tau_i $ - default time of name $ i $ 
\item $  \tau^j  $ - time of the $ j$-th default in the portfolio 
\item $ [T^{(1)}, \ldots, T^{(M)} ] $  - reference maturities\footnote{
The choice of the grid $ [T^{(1)}, \ldots, T^{(M)} ] $ is determined by
the desired resolution in calibration to single name default 
probabilities.}. We set $ T^{(0)}  = 0 $.
\end{itemize}
Next, we introduce our modeling primitives.
For each interval $ m = 1, \ldots, M $, we consider the following matrix
\beq
\label{firstForm}
P_{ij}^{(m)} (t)= P \left[  \left\{ \tau^j = \tau_i \right\} \bigcap \left\{ T^{(m-1)} \leq 
\tau_i \leq T^{(m)} \right\} | \mathcal{F}_t \right].
\eeq
In words, this is the joint probability that the $ i $-th name is the $ j$-th defaulter, 
and this event happens in the interval $ [T^{(m-1)},T^{(m)} ] $, conditional on the 
currently available information $ \mathcal{F}_t $. 
(Modeling of filtration $ \mathcal{F}_t $ will be discussed below.) Note that we assume instantaneous multiple default events are not possible.
Clearly, we can equivalently write (\ref{firstForm}) as 
\beq
\label{secondForm}
P_{ij}^{(m)}(t) = P \left[ \left. \left\{ \tau_i  = \tau^j  \right\} \bigcap \left\{ T^{(m-1)} \leq 
\tau^j  \leq T^{(m)} \right\} \right| \mathcal{F}_t \right].
\eeq
In what follows, we will occasionally refer to (\ref{firstForm}) or (\ref{secondForm}) as Top-Down
default time matrices, or TD-matrices for short.
The definitions (\ref{firstForm}), (\ref{secondForm}) are inspired by the formalism
used in the reliability theory and competing risks 
models\footnote{We want to thank Philipp Sch{\"o}nbucher for showing one of us (I.H.) 
formulas nearly identical to (\ref{firstForm}), (\ref{secondForm}).  While the formalism based
on joint probabilities (\ref{firstForm}) turns out to be equivalent to one 
previously employed in 
\cite{IHpresentation} which instead uses conditional probabilities (see below), 
we find it convenient to start the exposition by introducing them first.}, where the object of inference
is the joint probability $ p(T, O) $ of the event $ T $ (e.g. a first failure in a system) and
the type $ O $ of risk among a set of alternatives $ O_1, O_2, \ldots $ that could cause 
the failure. In our setting, identities of defaulted names serve as risk types 
$ O_i $, with an additional rule 
that they cannot repeat in the default history of a 
portfolio\footnote{This rule is easy to impose in a Monte Carlo setting, see
below.}.

Both the single name and portfolio probabilities are obtained by marginalization:
\bea
\label{marginals}
\sum_{i} P_{ij}^{(m)} (t) &=&  P \left[ \left. T^{(m-1)} \leq \tau^j  \leq T^{(m)} 
 \right| \mathcal{F}_t \right], \nonumber \\
\sum_{j} P_{ij}^{(m)}(t) &=&  P \left[ \left. T^{(m-1)} \leq \tau_i  \leq T^{(m)} 
 \right| \mathcal{F}_t \right],      
\eea
where we used (\ref{secondForm}) and (\ref{firstForm}), respectively.
Both the ``top-down" and ``bottom-up" forward probabilities
entering the RHS of Eqs.(\ref{marginals}) can be easily calculated
as follows:
\bea
\label{RHSmarginal}
P \left[ \left. T^{(m-1)} \leq \tau^j  \leq T^{(m)}  \right| \mathcal{F}_t \right]  
&=&  w_t^{(m)}(j) - w_t^{(m-1)}(j), \nonumber \\
P \left[ \left. T^{(m-1)} \leq \tau_i  \leq T^{(m)}  \right| \mathcal{F}_t \right] 
&=& Q_{i,t}^{(m)} - Q_{i,t}^{(m-1)},
\eea
where $ w_t^{(m)}(j) $ is the tail probability of having at least 
$ j $ defaults 
in the portfolio 
by time $ T^{(m)} $ (note that $ w_0^{(0)}(j) =  \delta_{j0} $), 
and $ Q_{i,t}^{(m)} $ is the cumulative default probability of the 
$ i $-th name by time $ T^{(m)}$. Note that we explicitly show dependence on $ t $ for
both $ w_t^{(m)}(j) $ and $ Q_{i,t}^{(m)} $ to emphasize their dependence on the 
filtration $ \mathcal{F}_t $.

\subsection{Dynamics of TD matrices}

The TD matrices are dynamic because we condition on the currently available information $\mathcal{F}_t$. In fact, from their definition it is clear that for any $i,j,m$, $P_{ij}^{(m)}(t)$ is a martingale in the filtration $(\mathcal{F}_t)_{t\ge0}$. In the remainder of this Section, as well as Sections \ref{SectionThinning} and \ref{SectionSens}, we examine the TD matrices as seen at time zero, i.e. conditional on $\mathcal{F}_0$. In Section \ref{SectionDynamics}, we show how the TD matrices are updated conditional on the information in $(\mathcal{F}_t)_{t\ge0}$. This must be done in a way that preserves 
the equations (\ref{marginals}) and (\ref{RHSmarginal}) as well as the martingale property.

The primary filtration $(\mathcal{F}_t)_{t\ge0}$ is constructed from the following definitions:
\begin{itemize}
	\item $(\mathcal{G}_t)_{t\ge0}$ is the natural filtration of the top-down 
default-counting process: $\mathcal{G}_t = \sigma( \{\tau^j<t \}; j=1,\ldots,N )$
	\item $(\mathcal{I}_t)_{t\ge0}$ is the filtration generated by the defaulters' identities: $\mathcal{I}_t = \sigma( \{\tau^j=\tau_i, \tau^j<t\}; i,j=1,\ldots,N)$
	(this will be obtained by simulation, see Sects. 2.3 and 5.1 below)
	\item $(\mathcal{H}_t)_{t\ge0}$ is a background filtration containing external market information, and possibly generated by several ``information processes'' 
	\item The filtrations $(\mathcal{G}_t)_{t\ge0}$, $(\mathcal{I}_t)_{t\ge0}$, and $(\mathcal{H}_t)_{t\ge0}$ are assumed to be independent, and the filtration $(\mathcal{F}_t)_{t\ge0}$ is defined by
		\bea
			\mathcal{F}_t = \mathcal{G}_t \vee \mathcal{I}_t \vee \mathcal{H}_t.
		\eea
\end{itemize}
According to these definitions, the ordered default times $\tau^j$ are $\mathcal{G}$-stopping times, but the single name default times $\tau_i$ are typically not. However, both $\tau^j$ and $\tau_i$ are $\mathcal{F}$-stopping times.

The market filtration $\mathcal{H}$ is similar to that introduced by Brody, Hughston and Macrina (BHM)
\cite{BHM}, which they use to introduce bond price dynamics to their model. In their paper, the filtration is generated by a Brownian bridge. By including this information in our overall filtration $\mathcal{F}$, we use similar techniques to enrich the dynamics of $P_{ij}^{(m)}(t)$ (see Section \ref{SectionDynamics}). In particular, by enlarging the filtration of interest, we are able to de-correlate (to a certain extent) the individual forward default probabilities from the index's; without it, they would be perfectly correlated.

\subsection{Relation to conditional forward matrices}

Note that for $ m = 1 $, our TD matrix $ P_{ij}^{(1)} $ is simply related to  
a time-independent matrix $ p_{ij} $ introduced by Ding, Giesecke and Tomecek 
\cite{Giesecke3} 
as a matrix of {\it conditional} 
probabilities, provided we  restrict it to
the time horizon given by $ T^{(1)} $, $  p_{ij} \rightarrow  p_{ij}^{(1)}(t)  
\equiv P \left[ \tau^j = \tau_i | \tau^j < T^{(1)}, \mathcal{F}_t \right] $. 
Indeed, in this case we obtain
\bea
\label{firstHorizon}
P_{ij}^{(1)}(t) &=& P \left[ \left\{ \tau^j = \tau_i \right\} \bigcap 
\left\{ 0 \leq \tau^j \leq T^{(1)} \right\} | \mathcal{F}_t \right] \\
&=&  P \left[  \tau^j = \tau_i  |  
\tau^j \leq T^{(1)}, \mathcal{F}_t  \right] \, 
P \left[ \tau^j \leq T^{(1)} | \mathcal{F}_t \right] 
\equiv  
p_{ij}^{(1)}(t) w^{(1)}_t(j). \nonumber
\eea
On the other hand, for a forward interval $ [ T^{(m-1)},T^{(m)}] $, we can 
similarly introduce a {\it conditional} forward TD-matrix
\beq
\label{condTD} 
p_{ij}^{(m)}(t) \equiv
P \left[ \tau^j = \tau_i | T^{(m-1)} \leq \tau^j \leq T^{(m)}, \mathcal{F}_t \right]. 
\eeq 
Using (\ref{RHSmarginal}) leads to the relationship between the conditional 
and joint TD-matrices
\beq
\label{equivalence}
P_{ij}^{(m)}(t) = p_{ij}^{(m)}(t)[w^{(m)}_t(j) - w^{(m-1)}_t(j) ].
\eeq

Applying (\ref{marginals}) leads to the marginal constraints for the conditional $p$-matrices:
\bea
\label{condConstraints}
\sum_{i} p_{ij}^{(m)}(t)  &=&  1, \nonumber \\
\sum_{j} p_{ij}^{(m)}(t) \left[ w^{(m)}_t(j) -
w^{(m-1)}_t(j) \right] &=&  
Q_{i,t}^{(m)} - Q_{i,t}^{(m -1)}.  
\eea
These are exactly the formulas presented earlier in \cite{IHpresentation}.

A comment is in order here. Eq.(\ref{equivalence}) demonstrates equivalence 
of the descriptions based on the joint and conditional forward TD-matrices
in the sense that as long as tail probabilities $ \{ w_j \} $ are known 
from the ``top'' part of the model, conditional TD-matrices are known once 
the joint matrices are known, and vice versa. However, as will be discussed below,
the joint TD-matrices are more convenient to work with when calculating 
sensitivities in the present formalism.

\subsection{Relation to intensity-based formulation}

For some applications, an equivalent formalism based on intensities, rather than 
probability matrices, may be preferred. To this end, we follow Giesecke and Goldberg
\cite{Giesecke1} and Ding, Giesecke and Tomecek \cite{Giesecke3} 
(see at the end of this section for a discussion) and introduce the 
so-called Z-factors as the conditional probability that name $ i $ is 
the next defaulter given an imminent default in the interval $ [t, t+dt] $:
\beq
\label{Z1}
Z_t^i = \sum_{n=1}^{N} P \left[ \tau^n = \tau_i | \tau^n \leq t + dt,  \mathcal{F}_t
\right] 1_{\{\tau^{n-1} \leq t < \tau^n\}},
\eeq
Note that (\ref{Z1}) yields the following relation between the single name and portfolio
intensities
\beq
\label{Z2}
\lambda_t^i = Z_t^i \lambda_t^p \; , \; \mbox{ where }\; \; \sum_{i=1}^N Z_t^i = 1.
\eeq
As shown by Giesecke and Goldberg, the single name 
default probability is given by the following relation:
\beq
\label{GGformula}
P \left[ t < \tau_i \leq T | \mathcal{F}_t \right] = \int_{t}^{T} \mathbb{E} 
\left[ \lambda_s^i | \mathcal{F}_t \right] ds =
\int_{t}^{T} \mathbb{E} 
\left[ Z_s^i \lambda_s^p | \mathcal{F}_t \right] ds.
\eeq
In general, one can take the $Z$-factors to be any $\mathcal{F}$-adapted stochastic processes satisfying (\ref{Z2}). 
To establish the relation with the probability-based formalism, we note 
that in our case of piecewise constant probabilities we can use the conditional 
forward TD matrices (\ref{condTD}) in Eq. (\ref{Z1}). This gives rise to 
the following ansatz for the $ Z$-factors:
\beq
\label{ansatz}
Z_t^i = \sum_{n=1}^N \left[ p_{in}^{(1)} 1_{ \{ \tau^{n-1} \leq t < \tau^{n} \} } 
1_{ \{ t \leq T_1 \} } + 
 p_{in}^{(2)} 1_{ \{ \tau^{n-1} \leq t < \tau^{n} \} } 
1_{ \{ T_1 < t \leq T_2 \} } + \ldots \right].
\eeq
Substituting this into the Giesecke-Goldberg formula (\ref{GGformula}), we obtain 
for the default probability at the first maturity: 
\bea
\label{Q1}
Q_{T_1}^i & \equiv & P \left[ 0 < \tau^i \leq T_1 \right] = 
\int_{t}^{T_1} \mathbb{E} \left[ Z_s^i 
\lambda_{s}^p \right] ds 
= \sum_{j=1}^{N} p_{ij}^{(1)} \int_{0}^{T_1} \mathbb{E} \left[ 
\lambda_{s}^p 1_{ \{ N_s = j-1 \} } \right] ds \nonumber \\ 
&=& \sum_{j=1}^{n} p_{ij}^{(1)} w_{0,T_1}(j),
\eea
and for the second maturity
\bea
\label{Q2}
Q_{T_2}^{i}  & \equiv &  P \left[ 0 < \tau^i \leq T_2 \right] = 
\int_{0}^{T_2} \mathbb{E} \left[ Z_s^i \lambda_{s}^p \right] ds 
= \int_{0}^{T_1} E \left[ Z_s^i \lambda_{s}^p \right] ds 
+ \int_{T_1}^{T_2} E \left[ Z_s^i \lambda_{s}^p \right] ds \nonumber \\
&=& Q_{T_1} + \sum_{j=1}^{n} P_{ij}^{(2)} \left( w_{0,T_2}(j) 
- w_{0,T_1}(j) \right) 
\equiv  Q_{T_1} + \sum_{j=1}^{n} P_{ij}^{(2)} w_{T_1,T_2}(j).
\eea
In the general case, we reproduce (\ref{condConstraints}).

A comment is in order here. Recently, Bielecki, Cr{\'e}pey and 
Jeanblanc (BCJ) \cite{BCJ} have pointed out that actual defaults in a credit portfolio 
are generally not stopping times under the information set $ \mathcal{G}_t  $ generated by the history of 
a pure "top"-dynamics. As a consequence, hazard rates obtained with a "classical"
RT procedure (which only uses filtration $ \mathcal{G}_t $) do not drop to zero
upon defaults, and thus cannot be mapped on the actual CDS spreads that do.
Our approach to handle this potential issue with a RT approach is to 
{\it extend the model filtration} $ \mathcal{F}_t $ (see Sect. 2.1) by including 
defaulter identities. Fortunately, in a Monte Carlo setting which we envision for 
applications of our framework, this can be easily done by {\it simulating} identities      
of defaulters upon each portfolio-level default (see  below in Sect. 5.1). 
Note that by doing this, we effectively turn the point process of losses in the top model
into a {\it marked} point process, where the marker provides the defaulter's 
identity.  Clearly, we should prevent multiple defaults of the same 
name, as well as ensure that
default intensity of a name drops to zero once it defaulted. 
We achieve both these goals
simultaneously within our Monte Carlo scheme as discussed further in Sect. 5.1.
      
\section{Thinning by bootstrap and iterative scaling}\label{SectionThinning}

In this section we present an algorithm that enables calculation of conditional
TD-matrices 
$ p_{ij}^{(m)}(t) $ (or, equivalently, joint TD matrices 
$ P_{ij}^{(m)} $) as seen today, at $ t = 0 $\footnote{
Some of the results of this section have been 
previously reported in \cite{IHpresentation}.}.

As the calibration scheme of TD matrices $ p_{ij}^{(m)}(0) $ is 
identical for all periods $ [T_{m-1},T_m] $, $ m = 1,\ldots, M $, in this
section we assume some fixed value of $ m $, and denote
$ p_{ij} \equiv p_{ij}^{(m)}(0) $,
$ w_j  \equiv w_0^{(m)} (j) - w_0^{(m-1)} (j) $, and 
$ Q_i \equiv Q_{i,0}^{(m)} - Q_{i,0}^{(m-1)} $.

Using this notation, our problem is to find a matrix $ p_{ij} $ that satisfies
the following row and column constraints:
\bea
\label{constraints1}
\sum_{j=1}^{N} p_{ij} w_j &=& Q_i \, , \; \; i = 1, \ldots, N, \\
\label{constraints2}
\sum_{i=1}^{N} p_{ij} &=& 1 \, , \; \; j = 1, \ldots, N. 
\eea
Note that this problem is ill-posed (and 
under-determined), as we 
have $ N^2 $ unknowns but only $ 2 N $ constraints, therefore
it can have an infinite number of solutions, or no solution at all, which 
happens when the constraints are contradictory. Before presenting the solution,
we therefore want to analyze necessary conditions for the existence of a
solution.
 
\subsection{Consistency condition}

If we sum Eq. (\ref{constraints1}) over i, we obtain
\beq
\label{consistency}
\sum_{j=1}^{N} w_j = \sum_{i=1}^{N} Q_i.
\eeq
Note that the RHS of this equation $ \sum_{i=1}^{N} Q_i = \la N_T \ra^{(CDS)} $ is
the expected number of defaults according to single-name (CDS) data, while the 
LHS gives the expected number of defaults according to the top-down model:
\beq
\label{lhs}
\sum_{j=1}^{N} w_j = \sum_{j=1}^{N} \sum_{n \geq j}^{N} p_n = \sum_{n=1}^{N} n p_n 
= \la N_T \ra^{(model)}.
\eeq
We thus have the following consistency condition:
\beq
\label{NN}
\la N_T \ra^{(model)} =  \la N_T \ra^{(CDS)} = \la N_T \ra.
\eeq
Unless (\ref{NN}) is satisfied, no set of top-down matrices
can match both the ``top'' and ``down'' data. Note that the standard 
procedure of basis adjustment ensures that the theoretical formula for the index 
par spread $ S_{idx} $ in terms of CDS par spreads $ S_i $ and their risky
durations $ PV01_i $: 
\beq
\label{indBas}
S_{idx} = \frac{\sum_{i} S_i PV01_i}{\sum_{i} PV01_i },
\eeq
holds by adding a uniform or proportional tweak to all spreads $ S_i $ in the index
portfolio. This procedure does {\it not} guarantee that the constraint (\ref{NN})
is met, and indeed one typically finds a few percent difference between
$ \la N \ra^{(model)} $ and $ \la N \ra^{(CDS)} $ even after the basis adjustment 
is made. While alternative schemes of basis adjustment where (\ref{NN}) would be 
enforced are certainly feasible, in this paper we choose to further adjust 
(basis-adjusted) single name default probabilities so that
(\ref{NN}) holds.

\subsection{Choice of the initial guess}

Obviously, a solution of an ill-posed problem defined by (\ref{constraints1}) and (\ref{constraints2}), if 
it exists, should generally depend on an initial guess (a ``prior'') $ q_{ij} $ 
for the TD matrix $ p_{ij} $.
Here we present a few possible specifications of the prior matrix.

One possible choice is a factorized prior $ q_{ij} = q_i k_j $. 
Consider the first three steps of the IS algorithm with this prior:
\bea
\label{firstThree}
p_{ij}^{(1)} &=& q_i k_j \frac{Q_{i}}{q_i \sum_{j} k_j w_j } = \frac{k_j
Q_i }{ \sum_{j} k_j w_j }, \nonumber \\
p_{ij}^{(2)} &=& \frac{p_{ij}^{(1)}}{\sum_{i} p_{ij}^{(1)}} = 
 \frac{k_j
Q_i }{ \sum_{j} k_j w_j } \frac{ \sum_{j} k_j w_j }{k_j \sum_{i} Q_i } 
= \frac{Q_i}{\sum_{i} Q_i } = \frac{Q_i}{\la N_T \ra }, \\
p_{ij}^{(3)} &=& p_{ij}^{(2)} \frac{Q_{i}}{\sum_{j} p_{ij}^{(2)} w_j } = 
 \frac{Q_i}{\la N_T \ra }, \nonumber
\eea
so that the algorithm converges to the solution 
$ p_{ij} = Q_i / \la N_T \ra $, independently of the initial guess. 
This solution suggests that the relative riskiness of a name stays the same
(i.e. independent of $ j $) as defaults arrive. However, such a behavior 
looks unreasonable on ``physical'' grounds, as riskier names
are expected to default earlier than less risky ones. We view this as
an evidence against using factorized priors.

\begin{figure}[ht]
\begin{center}
\includegraphics[width=100mm,height=60mm]{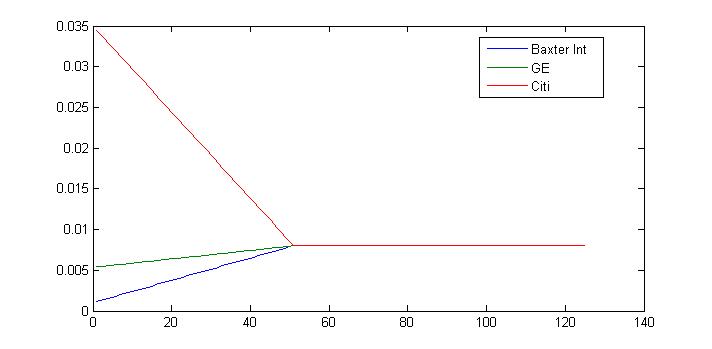}
\caption{Three rows of the prior matrix $ q_{ij}^{(1)} $ corresponding
to low, moderate and high spreads. } \label{fig:priorLinear}
\end{center}
\end{figure}

A simple and reasonable alternative to factorized priors that conforms to one's
intuition about the order of defaulters in the portfolio is obtained if
we assume a linear law for rows of conditional p-matrix, such that for 
a risky name the probability that it defaults first, second, etc. will  
(linearly) decrease, while for tighter names it will increase.
We can further
assume that once some sufficiently high default level $ \bar{n} $ 
is reached, the conditional
TD-matrix $ p_{ij}$ becomes uniform across names, so that 
$ \bar{n} $ can be referred to as a ``uniformization bound''. 
This is summarized in the 
following ansatz for the prior matrix:
\bea
\label{prior}
q_{ij}  
= \left\{ \begin{array}{ll}
q_{i1} + \alpha_i j 
& \mbox{$i = 1, \ldots, N, j \leq \bar{n} $} \\
\frac{1}{N} & \mbox{$i = 1, \ldots, N, j > \bar{n},  $}
\end{array}
\right.
\eea
where
\[
 \alpha_i = \frac{1}{\bar{n}} \left( \frac{1}{N} - q_{i1} \right). 
\]
This ansatz parametrizes
the p-matrix in terms of its first column. The latter can be chosen according
to current values of CDS spreads. Note that $ \sum_{i} q_{ij} = 1 $ as long 
as $ \sum_{i} q_{i1} = 1 $. In Fig.~\ref{fig:priorLinear} we 
show three rows in such a prior matrix for the first maturity interval, 
corresponding to names with low, 
moderate and high spreads (resp., Baxter International, GE and Citigroup).

The ansatz (\ref{prior}) is certainly not the only possible model for the prior.
Another method to construct the prior will be presented below after we discuss
the single name sensitivities calculation. 

\subsection{Iterative scaling (IS) algorithm}

\begin{figure}[ht]
\begin{center}
\includegraphics[width=50mm,height=50mm]{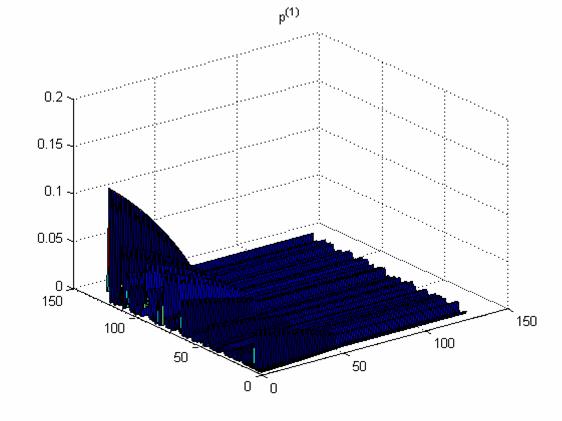}
\includegraphics[width=50mm,height=50mm]{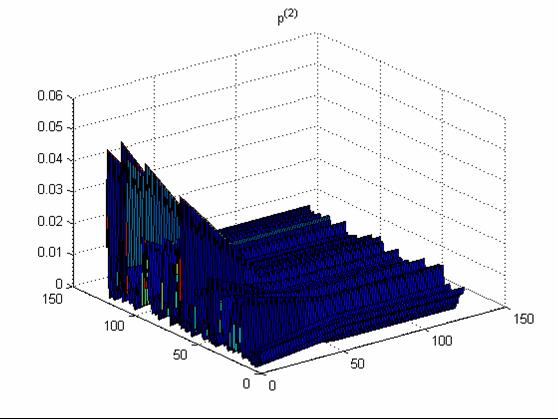}
\includegraphics[width=50mm,height=50mm]{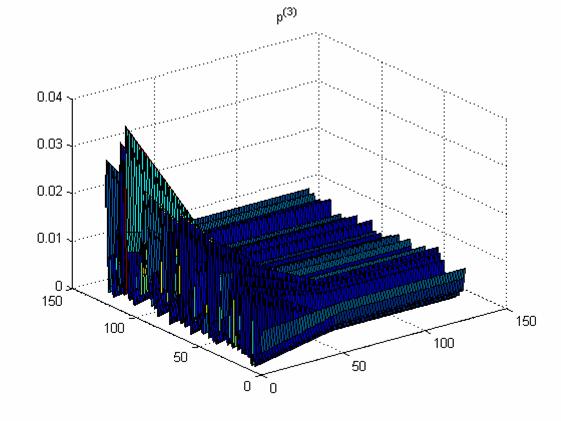}
\caption{Calibrated thinning matrices $ p_{ij}^{(1)}, p_{ij}^{(2)},
p_{ij}^{(3)} $ for intervals $ [0,3Y], \, [3Y,5Y], \, 
[5Y,7Y] $ for CDX IG8 data on 03/03/2008, 
obtained with linear prior (\ref{prior}). (The fourth matrix is similar and 
is not shown here to save space.)} \label{fig:resultLinearPrior}
\end{center}
\end{figure}

\begin{figure}[ht]
\begin{center}
\includegraphics[width=100mm,height=60mm]{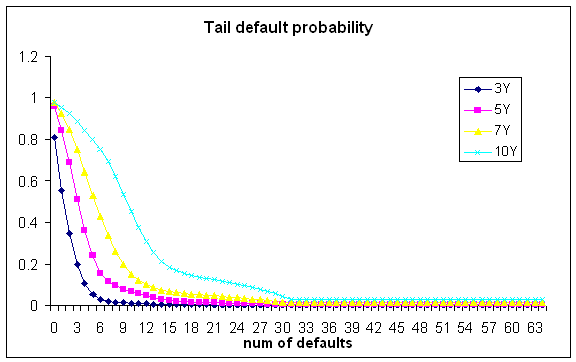}
\caption{Tail probabilities for horizons of 3,5,7 and 10Y for CDX IG8 data on 03/03/2008
calculated with the BSLP model. } \label{fig:TailProbs}
\end{center}
\end{figure}

We assume that an initial guess (a ``prior'') 
$ q_{ij} \equiv p_{ij}^{(0)} $ for the solution is available using
e.g. a construction just presented (see below for an alternative choice).
We then use the iterative scaling algorithm\footnote{The method was 
originally developed
in 1937 by Kruithof to estimate telephone traffic matrices.
For more information on the IS method and relation to 
information theory, see \cite{Csiszar} and also below.} 
to find the matrix $ p_{ij} $. With 
this 
method, the matrix is updated iteratively $ p^{(0)} \rightarrow p^{(1)}
\cdots \rightarrow p^{(k)} \rightarrow p^{(k+1)} \cdots $ according to the 
following scheme:
\bea
\label{bMatEl1}
p_{ij}^{(k+1)}  
= \left\{ \begin{array}{ll}
p_{ij}^{(k)} \frac{Q_i}{\sum_{j} p_{ij}^{(k)} w_j} 
& \mbox{for $k $ odd} \\
 \frac{p_{ij}^{(k)}}{\sum_{i} p_{ij}^{(k)}} & \mbox{for $ k $ even.}
\end{array}
\right.
\eea
In other words, we alternatively rescale the matrix to enforce the row and column
constraints until convergence.
The equivalent scheme for joint TD-matrices $ P_{ij} $ reads
\bea
\label{bMatEl}
P_{ij}^{(k+1)}  
= \left\{ \begin{array}{ll}
P_{ij}^{(k)} \frac{Q_i}{\sum_{j} P_{ij}^{(k)}} 
& \mbox{for $k $ odd} \\
P_{ij}^{(k)} \frac{w_j}{\sum_{i} P_{ij}^{(k)}} & \mbox{for $ k $ even.}
\end{array}
\right.
\eea
We have tested the above algorithm on several datasets, and found in each case fast 
convergence (in less then 10 steps per maturity) to parameters matching single name 
CDS spreads with relative errors below 1\%. The whole calculation takes about 2-3 seconds
on a standard PC. An example of calibrated conditional thinning matrices is given in 
Fig.~\ref{fig:resultLinearPrior} for CDX IG8 data on 03/03/2008 (the 
same dataset will be used for all numerical examples in what follows).
Tail probabilities needed for this calculation are produced by our own version of a 
top-down model called BSLP (Bivariate Spread-Loss Portfolio) model \cite{AH}, however
any other top-down model can be used to this end. Tail probabilities produced by BSLP
by calibrating to the same dataset are shown in Fig.~\ref{fig:TailProbs}.
 
\subsection{Information-theoretic interpretation of IS algorithm}

If we rescale our matrices $ p_{ij} \rightarrow N p_{ij} $ and
$ q_{ij} \rightarrow N q_{ij} $ so that now $ \sum_{i,j} p_{ij} = 
\sum_{i,j} q_{ij} = 1 $, then $ \{ p_{ij} \}, \{ q_{ij} \} $ can
be thought of as two-dimensional probability distributions.     
As was shown by Ireland and Kullback \cite{IK} and Csisz{\'a}r \cite{Cs_1975} 
(see also \cite{Csiszar}), the IS algorithm  
can be interpreted as an alternative minimization of the 
Kullback-Leibler divergence (relative entropy, see e.g. \cite{CT})
between these two measures:
\beq
\label{KL}
D[p||q] = \sum_{ij} p_{ij} \log \frac{p_{ij}}{q_{ij}}, 
\eeq
subject to constraints $ \sum_{j=1}^{N} p_{ij} w_j = \frac{1}{N} Q_i $ and 
$ \sum_{i=1}^{N} p_{ij} = 1/N $. The standard approach to this problem
uses the method of Lagrange multipliers to enforce the constraints,
leading to the following Lagrangian function
\beq
\label{Lagrangian}
\mathcal{L} = D[p||q] - \sum_{j} \lambda_j \left( \sum_{i} p_{ij} - 
\frac{1}{N} \right)
- \sum_{i} \xi_i \left( \sum_{j} p_{ij} w_j - \frac{Q_i}{N} \right),  
\eeq
which can be solved using convex duality \cite{CT}, with dimensionality of 
the problem being the number of Lagrange multipliers $ \xi_i $, i.e. equal to
the number of constraints. The approach of Ireland and Kullback and Csisz{\'a}r
uses instead an {\it alternative} recursive minimization of the KL distance 
(\ref{KL}) where we alternate between enforcing the row constraints 
only at odd steps, and column constraints only at even steps.
This can be shown to precisely correspond to the alternating rescaling of the IS
algorithm. 
This method may be computationally less expensive than the direct minimization 
of (\ref{Lagrangian}), especially for large scale problems, and it allows one 
to prove convergence of the IS algorithm under certain technical conditions.
 
We note that the information-theoretic 
reformulation of the problem can be easily generalized to the case when 
one matches single name spreads approximately (e.g. in the least square sense) 
rather than exactly, and thus can be used to ensure stability of single name 
calibration. We will not pursue this approach further in this paper.
 
\section{Single name sensitivities} \label{SectionSens}

The TD-matrices calibrated as of today can be used for calculation of single name sensitivities, as proposed by 
Giesecke and co-authors \cite{Giesecke1,Giesecke2,Giesecke3}. Specifically, once
a calibrated set of TD-matrices is found, sensitivity to a given name $ i $ is 
calculated 
as follows. First we re-scale the $i$-th row in the unconditional TD-matrix to accommodate the 
new (bumped)
spread of the $i$-th name. Once this is done, summation of the new perturbed 
matrix over $ i $
produces a new set of bumped tail probabilities, which is 
equivalent to 
bumping tranche prices. The ratio of changes of the MTM of a given 
tranche to that of the underlying names is the name delta of the tranche.

Let us consider the calculation in more details, focusing  for simplicity on
the first maturity interval $ [0,T_1 ] $. Assume that we want to calculate
sensitivity of tranche prices to a given name $ i $. To this end,
we tweak only the $ i$-th row of the joint TD matrix $ P_{ij} = P_{ij}^{(1)}$.
We consider the simplest proportional tweak of all elements in the $ i $-th 
row\footnote{We choose a proportional tweak as another simple one-parametric 
scheme with an additive tweak $ \delta P_{ij} = \varepsilon $ would lead to 
nearly equal sensitivities for different names, see a comment after 
Eq.(\ref{SingleNameDelta}).}:
\beq
\label{propTweak}
\delta P_{ij} = \varepsilon P_{ij}.
\eeq
Using Eqs.(\ref{marginals}), we obtain the following tweaks of the default 
probability $ Q_i(T) $ and tail probabilities $ w_j $:
\bea
\label{dQdW}
\delta Q_i &=& \sum_{j=1}^{N} \delta P_{ij} = \varepsilon \sum_{j=1}^{N} P_{ij} = 
\varepsilon Q_i, \nonumber \\
\delta w_j &=& \delta P_{ij} = \varepsilon P_{ij}.
\eea
Note that as long as the rule (\ref{propTweak}) 
of tweak of $ P_{ij} $ is specified, the tweak of 
the conditional TD-matrix $ p_{ij} $ is not arbitrary but is rather fixed by the 
relation
\beq
\label{dP}
\delta P_{ij} = w_j \delta p_{ij} + p_{ij} \delta w_j.
\eeq
Substituting (\ref{propTweak}) and 
the second of Eqs.(\ref{dQdW}) and re-arranging, this yields
\beq
\label{dp}
\delta p_{ij} = \varepsilon p_{ij} \left( 1 - p_{ij} \right).
\eeq

Eqs.(\ref{dQdW}) express the sought-after ``duality'' between 
tweaks of single name 
default probability and the tail probability, which both stem from
a tweak of the joint TD-matrix $ P_{ij} $. It is exactly this duality that 
enables the whole calculation of single name sensitivities in our 
framework\footnote{We note that such a tying of spread shifts to particular 
shifts of tail probabilities (and hence correlation parameters) is akin to the local
volatility approach in equity derivative modeling.}.
What remains now for calculation 
of those single name sensitivities is to establish
a relation between changes of the market-to-market of name $ i $ and a tranche with 
changes $ \delta Q_i $ and $ \delta w_j $, respectively. 
 
In what follows, we establish approximate, rather than ``exact'', relations for 
single name sensitivities. Generalizations of formulae to follow to make them
``exact'' (i.e. include contributions of all cashflows) are straightforward, but lead
to bulky expressions and hence will be omitted here. Furthermore, as will be shown later
on, our approximate formulas can be inverted, leading to some interesting
insights (see Sect. \ref{SectionSens}.1).

We start with establishing relevant formulae for name $ i $.
Let us approximate a change of MTM, $ \delta MTM_{i} $, of CDS referencing the 
$ i$-th name by a change $ \delta DL_i $ of its default leg. (Note that this
approximation becomes exact if the CDS premium is paid upfront rather than as a 
running spread. Corrections to this approximation will be considered below.)  
In turn, $ \delta DL_i $ 
can be approximated by the (discounted) change in 
the expected loss on this name (i.e. the default probability times 
$ (1-R)/N$, assuming a unit portfolio notional). This yields 
\beq
\label{dMTMi}
\delta MTM_i \simeq \frac{1-R}{N} B(0,T) \delta Q_i(T),
\eeq
where $ B(0,T) $ is a discount factor to time $ T$\footnote{Corrections
to the RHS of this equation are $ O(r) $ where $ r $ is a short risk-free rate,
and thus can be neglected as long as $ r $ is small enough.}.

We can now use the same approximation as above 
for the MTM change of the tranche, so that the
latter is approximated by the (discounted) change of the tranche expected loss (EL).
We assume an ordered set of strikes $ K_1, \ldots, K_{N_{tr}} $ expressed as a fraction
of the total portfolio notional, with $ K_0 = 0 $ and $ K_{N_{tr}} = 1 $.
Let $ EL_k \equiv EL_{[K_{k-1}, K_k ]}
$ (where $ k = 1,\ldots, N_{tr} $) be the expected loss of the $k$-th tranche,
expressed as a percentage of the tranche 
notional, 
is defined as follows (here $ I = \frac{1}{N} $ is the CDS notional):
\beq
\label{ELdef}
EL_{k} = \frac{S_{k-1} - S_k}{K_{k} - K_{k-1}},
\eeq
where
\beq
S_k = \sum_{n=1}^{N} \left( (1-R) n I - K_k N I\right)^{+} \rho_n,
\eeq
and $ \rho_n $ is the probability of having $n$ defaults by a given time horizon, 
which can be expressed in terms of the tail probabilities $ \{ w_n \} $ defined in Sect.\ref{SectionTD}:
\bea
\label{W2p}
\rho_n = \left\{ \begin{array}{ll}
w_n - w_{n+1} & \mbox{$ n = 1, \ldots, N-1 $} \\
w_N & \mbox{$ n = N $.}
\end{array}
\right.
\eea
After some algebra, we obtain for $ S_k$
\[
S_k = \frac{1-R}{N} 1_{(\hat{K}_k < 1)} \left( \sum_{ \lfloor \hat{K}_k N
\rfloor }^{N} w_n + \left( \lfloor \hat{K}_k N \rfloor - \hat{K}_k N \right) 
w_{\lfloor \hat{K}_k N \rfloor} \right) 
\simeq  \frac{1-R}{N} 1_{(\hat{K}_k < 1)} \sum_{n= \lfloor \hat{K} N
\rfloor }^{N} w_n, 
\]
where $ \hat{K}_k \equiv K_k/(1-R) $ and $ \lfloor x \rfloor $ stands for 
the smallest integer equal or larger than $ x $.
For the tranche expected loss with $ \hat{K}_{k-1} , \hat{K}_k < 1 $ we 
therefore have
\beq
\label{trancheEL}
EL_{ k } 
 \simeq  \frac{1-R}{K_k - K_{k-1}} 
\frac{1}{N} \sum_{ n = \lfloor \hat{K}_{k-1} N \rfloor}^{ 
\lfloor \hat{K}_k N \rfloor} w_n.
\eeq

Using (\ref{trancheEL}), (\ref{dMTMi}) and (\ref{dQdW}), we arrive at the following 
approximate formula for the single name delta of the $ k$-th tranche 
in the random thinning framework:
\bea
\label{SingleNameDelta}
\Delta_i^{k} &\equiv& \frac{(\delta MTM)_{tranche}^k}{(\delta MTM)_i} \simeq
\frac{1}{K_k - K_{k-1}} \frac{1}{N} \frac{\sum_{n = \lfloor \hat{K}_{k-1} 
N \rfloor }^{\lfloor \hat{K}_k N \rfloor} \delta P_{in} }{
\sum_{n = 1}^{N} \delta P_{in} } \nonumber\\
&=&
\frac{1}{K_k - K_{k-1}} \frac{1}{N} \frac{\sum_{n = \lfloor \hat{K}_{k-1} 
N \rfloor }^{\lfloor \hat{K}_k N \rfloor} w_n  p_{in} }{
\sum_{n = 1}^{N} w_n  p_{in} }.
\eea
Note that this formula readily demonstrates that tweaks 
$ \delta P_{in} = \varepsilon w_n p_{in} $ should
be different for different names, as otherwise our approximation would produce
the same deltas for all names.

We can also calculate the index delta from all single name deltas as follows. 
If we tweak all single names at once using a 
proportional tweak with the same $ \varepsilon $ for all names, then 
the change of tranche MTM is 
\beq
\label{dmtm_tr} 
(\delta MTM)_{tranche}^k = \sum_{i} \Delta_i^k (\delta MTM)_i, 
\eeq
while on the other hand by definition it is equal to 
\beq
\label{dmtm_idx}
(\delta MTM)_{tranche}^k = \Delta_{idx}^k (\delta MTM)_{idx}.
\eeq
Comparing these two formulas and taking into account the relation
\beq
\label{dMTMidx} 
(\delta MTM)_{idx} = \sum_{i} \frac{1}{N} (\delta MTM)_i = \varepsilon 
(1-R) B(0,T) \sum_{i=1}^{N} \sum_{n=1}^{N} w_n  p_{in},  
\eeq
we obtain two relations for the index delta. The first one is a ``bottom-up'' 
formula expressing the index delta as a weighted average of single name deltas:
\beq
\label{IdxDelta_bu}
\Delta_{idx}^k = N \frac{\sum_{i=1}^{N} \Delta_i^k (\delta MTM)_{i}}{
\sum_{i=1}^{N} (\delta MTM)_{i}},
\eeq
and the second one is a ``top-down'' relation 
\beq
\label{IdxDelta_td}
\Delta_{idx}^k = \frac{(\delta MTM)_{tranche}^k}{(\delta MTM)_{idx}}
 \simeq \frac{1}{K_k - K_{k-1}} \frac{
\sum_{i=1}^{N} \sum_{n = \lfloor \hat{K}_{k-1} N \rfloor}^{
\lfloor \hat{K}_k N \rfloor} w_n  p_{in}}{
\sum_{i=1}^{N} \sum_{n=1}^{N} w_n  p_{in}}.
\eeq
Note that approximate deltas
calculated according to (\ref{IdxDelta_td}), and 
rescaled by the tranche widths $ K_k - K_{k-1} $, 
sum up to 1 across all 
tranches in the capital structure, while single name 
deltas sum up to $ 1/N $. Also note the following simple relation between the index 
delta and single name deltas:
\beq
\label{idxDelta}
\Delta_{idx}^k \simeq \sum_{i} \Delta_i^k,
\eeq
which follows from (\ref{IdxDelta_bu}) 
as long as all spreads in the index portfolio are approximately 
equal and tweaked in the same way (i.e. using the same value of 
$ \varepsilon $). 

We would like to note that the accuracy of approximate formulas 
(\ref{SingleNameDelta}) and (\ref{idxDelta}) can be improved without adding 
complexity. This is done by assuming a continuous coupon, and approximating the 
premium legs of a CDS and a 
tranche using the a constant riskless rate  $ r $ and a
constant hazard rate $ h_i $ so that the
survival probability of name $ i $ reads  
\beq
\label{CDSdefProb}
Q_i(T) = e^{ - h_i T}.
\eeq
and a similar formula for the survival probability of a tranche. 
This yields the following value for the premium leg of a tranche
\bea
\label{PLff}
PL_{[K_d,K_u]} &=& s \int_{0}^{T} dt e^{-r t} e^{-ht} = \frac{1}{r+h}\left(1 - 
e^{-(r+h)T} \right) \simeq s T \left( 1 - \frac{1}{2} T (r+h) \right)
\nonumber \\
&\simeq & s T \left( 1 - \frac{1}{2}r T   - \frac{1}{2} 
EL_{[K_d,K_u]} \right),
\eea
where $h$ is the continuous coupon rate. 
This produces the following refinement to the approximate relation 
(\ref{SingleNameDelta}):
\beq
\label{SingleNameDelta2}
\Delta_i^k \simeq 
\frac{1}{K_k - K_{k-1}} \frac{1}{N} \frac{B(0,T) + \frac{1}{2} s_{tr}^k T}{
B(0,T) + \frac{1}{2}s_i T}
\frac{\sum_{n = \lfloor \hat{K}_{k-1} N  \rfloor }^{
\lfloor \hat{K}_k  N \rfloor} w_n  p_{in} }{
\sum_{n = 1}^{N} w_n p_{in} } . 
\eeq
Numerical experiments indicate that accuracy of the approximate formula 
(\ref{SingleNameDelta2}) in comparison with an ``exact'' expression that includes
contributions of all cashflows is not worse than 10\%.

The results of single name sensitivities calculated using (\ref{SingleNameDelta2}) 
are given in 
Tables ~\ref{Baxter}, ~\ref{GE} and ~\ref{Sprint} for three 
names representing low, moderate and high spread names, respectively.
\begin{table}
\vspace{5mm}
\begin{tabular}{|c|c|c|c|c|c|r|}
\hline
Tranche & 0-3\%  & 3-7\%  & 7-10\%  & 10-15\% & 15-30\% & 30-100\%     \\
\hline
BC delta & 0.0546 & 0.0086 & -0.0022 & -0.0034 & -0.0093 & 0.0109  \\
\hline
RT delta (prior 1) & 0.0521 & 0.0218 &  0.0152 & 0.0102 & 0.0085 &  0.0057 \\
\hline
RT delta (prior 2) & 0.0283  & 0.0035  & 0.0011  & 0.0005  & 0.0009 &  0.0101 \\
\hline
\end{tabular}
\caption{Single name sensitivities for a low spread name (Baxter).}
\label{Baxter}
\end{table}
\begin{table}
\vspace{5mm}
\begin{tabular}{|c|c|c|c|c|c|r|}
\hline
Tranche & 0-3\%  & 3-7\%  & 7-10\%  & 10-15\% & 15-30\% & 30-100\%     \\
\hline
BC delta &  0.0900 &  0.0314  & 0.0153  & 0.0136 &  0.0172 &  0.0006 \\
\hline
RT delta (prior 1) & 0.1180 & 0.0334 &  0.0175 & 0.0097 &  0.0056 &  0.0034 \\
\hline
RT delta (prior 2) & 0.1025  & 0.0345 &  0.0212 &  0.0159 & 0.0156 &  0.0014\\
\hline
\end{tabular}
\caption{Single name sensitivities for a medium spread name (GE).}
\label{GE}
\end{table}
\begin{table}
\vspace{5mm}
\begin{tabular}{|c|c|c|c|c|c|r|}
\hline
Tranche & 0-3\%  & 3-7\%  & 7-10\%  & 10-15\% & 15-30\% & 30-100\%     \\
\hline
BC delta           & 0.1914 & 0.0487 & 0.0116 & 0.0028 & 0.0001 & -4.75e-10  \\
\hline
RT delta (prior 1) & 0.1744 & 0.0433 & 0.0195 & 0.0093 & 0.0032 &  0.0013 \\
\hline
RT delta (prior 2) & 0.1725 & 0.0595 & 0.0316 & 0.0144 & 0.0021 &  0.0002\\
\hline
\end{tabular}
\caption{Single name sensitivities for a high spread name (Sprint).}
\label{Sprint}
\end{table}
A few comments are in order here in regard to these numbers. The first row in these 
tables labeled ``BC delta'' shows sensitivities calculated with the base correlation 
method. Negative entries in Table 1 (and 3) clearly show that the base correlation
methodology is ``wrong'' in the sense that for positive asset correlations, all 
single name sensitivities in a ``right'' model should be positive, not 
negative\footnote{This deficiency of the base correlation model is well known to 
practitioners.}. Nevertheless, in view of the absence of a market-standard alternative
to the base correlation methodology, we will keep base correlation sensitivities 
as a reference point for our RT scheme.  The second row (``RT delta (Prior I)'' in the 
tables refers to single name sensitivities calculated with the ``linear'' prior 
(\ref{prior}), while the third row gives the RT delta calculated with the base 
correlation prior which is explained below in Sect. \ref{SectionSens}.1. One sees that the RT method
produces numbers of the same order of magnitude as the base correlation model,
with largest differences being for mezzanine and senior tranches. 

At this point, a question that could be asked is which set of deltas is the ``right'' one. Generally, the ultimate answer to this question requires exploring the behavior of the P\&L distribution of the hedged position, where hedges are calculated according to the model. While we do not pursue such an analysis in this paper, we note that RT deltas are likely to be ``less wrong'' than base correlation deltas as they arise from a consistent, arbitrage-free model. For example, unlike the latter, positivity of RT deltas is guaranteed by construction. More on the comparison between RT sensitivities calculated with Prior I and Prior II will be said in Sect. \ref{SectionSens}.1, after the ``base correlation prior (``Prior II'') is introduced.

\subsection{``Base correlation'' prior}

Assume we are given some set of ``target'' single name sensitivities which we 
would like to match as closely as possible. These sensitivities can come from
any bottom-up model such as CreditMetrics, approximately calibrated to tranche
quotes of interest. Using the results of the last section, we can then invert
the relation (\ref{SingleNameDelta2}) or (\ref{SingleNameDelta})
and construct the $ p$-matrices such that these deltas are approximately matched.
We then use these $p$-matrices as priors that need to be corrected by the IS 
algorithm to match the single-name and portfolio data. The new ``true'' 
sensitivities are then calculated using the calibrated TD-matrices. 

\begin{figure}[ht]
\begin{center}
\includegraphics[width=100mm,height=60mm]{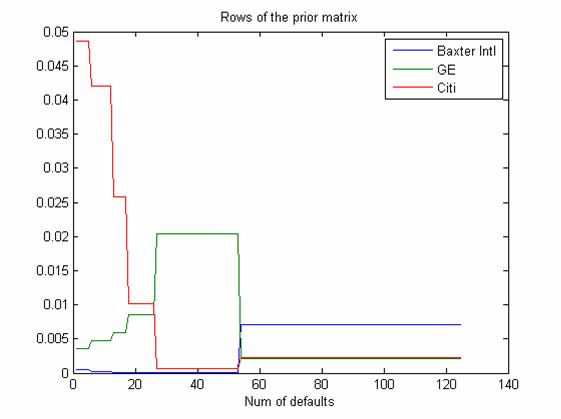}
\caption{Three rows of the prior matrix $ q_{ij}^{(1)} $ corresponding
to low, moderate and high spreads, with the ``base correlation''  
prior. } \label{fig:priorBC}
\end{center}
\end{figure}

\begin{figure}[ht]
\begin{center}
\includegraphics[width=50mm,height=50mm]{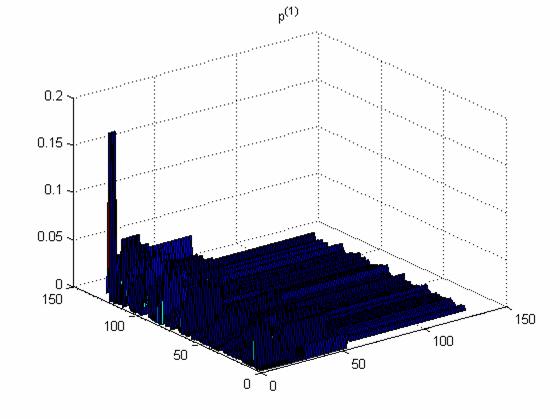}
\includegraphics[width=50mm,height=50mm]{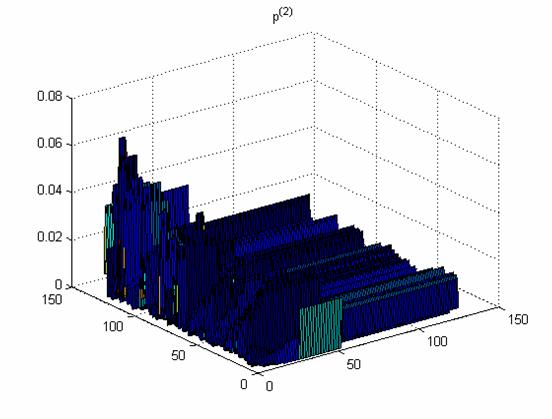}
\includegraphics[width=50mm,height=50mm]{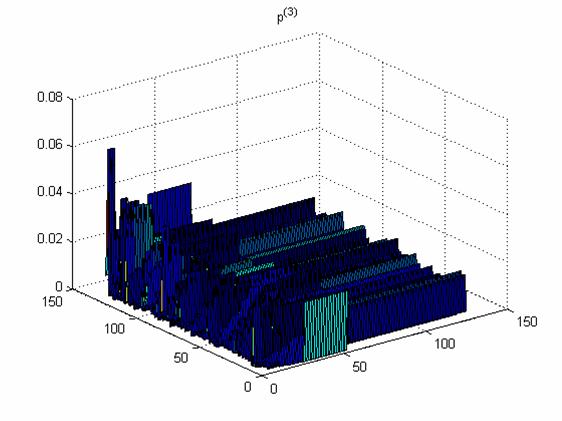}
\caption{Calibrated thinning matrices $ p_{ij}^{(1)}, p_{ij}^{(2)},
p_{ij}^{(3)} $ obtained with the ``base correlation prior''.} 
\label{fig:resultBCPrior}
\end{center}
\end{figure}

Note that each row in the $p$-matrix has $ N $ elements (e.g. N = 125 for CDX.NA.IG and
iTraxx portfolios), while for the same portfolios we have only 6 deltas per name.
This means that the problem is severely underdetermined.
To reduce the number of free parameters, we assume that for each $ i $ and $ m $, 
elements of $ p_{ij}^{(m)} $ are piecewise-constant between default counts 
$ j $ that correspond
to strikes of tranches in the calibration set\footnote{Note that these values of $ j$ 
are easily found as long as we assume a fixed recovery: $ j_k = \lfloor \hat{K}_k N 
 \rfloor $.}.
Thus, we set $ p_{in} =  \bar{p}_i^1 $ for $ j \in [1, j_1] $, 
$ p_{in} =  \bar{p}_i^2 $ for $ j \in [j_{1}+1, j_2] $, etc. 
Eq. (\ref{SingleNameDelta}) with a proportional tweak $ (\delta p)_{in} = \varepsilon
p_{in} $ produces the following formula for the delta of the $k$-th tranche
with respect to name $ i $ 
\bea
\label{SingleNameDelta3}
\Delta_i^k & \simeq  &
\frac{1}{K_k - K_{k-1}} \frac{1}{N} \frac{\sum_{n = \lfloor \hat{K}_{k-1} N \rfloor }^{
\lfloor \hat{K}_k N \rfloor} w_n p_{in} }{
\sum_{n = 1}^{N} w_n p_{in} } = 
\frac{1}{K_k - K_{k-1}} \frac{1}{N} \frac{\sum_{n = \lfloor \hat{K}_{k-1}N \rfloor }^{
\lfloor \hat{K}_k N \rfloor} w_n p_{in} }{Q_i(T_1) } \nonumber \\
&=& 
\frac{1}{K_k - K_{k-1}} \frac{1}{N} 
\frac{\bar{p}_i^k \sum_{n = \lfloor \hat{K}_{k-1}N \rfloor }^{
\lfloor \hat{K}_k N \rfloor} w_n }{Q_i(T_1) }  = 
\frac{1}{K_k - K_{k-1}} \frac{1}{N} 
\frac{\bar{p}_i^k  \bar{w}^k }{Q_i(T_1) },  
\eea
where $ \bar{w}^k \equiv \sum_{n = \lfloor \hat{K}_{k-1}N \rfloor }^{
\lfloor \hat{K}_k N \rfloor} w_n  $. Inverting this relation, we obtain
\beq
\label{pk}
\bar{p}_i^k = \Delta_i^k Q_i(T_1) \frac{ N (K_k - K_{k-1}) }{\bar{w}^k}.
\eeq
Similar to Eq. (\ref{SingleNameDelta2}), this relation can be improved by taking into
account non-vanishing spreads paid by a CDS and a tranche:
\beq
\label{pk1}
\bar{p}_i^k = \Delta_i^k Q_i(T_1) \frac{ N (K_k - K_{k-1}) }{\bar{w}^k}
\frac{B(0,T) + \frac{1}{2} s_{i} T}{
B(0,T) + \frac{1}{2}s_{tr}^k T}. 
\eeq
The result is shown in Fig.~\ref{fig:priorBC} for the same 
three names that were used for 
illustration of our ``linear'' prior matrices.
Resulting calibrated thinning matrices are shown in 
Fig.~\ref{fig:resultBCPrior}. They can be compared with the profiles 
displayed in Fig.~\ref{fig:priorBC}. One sees that the direction behavior 
is similar in both cases.

\begin{figure}[ht]
\begin{center}
\includegraphics[width=100mm,height=60mm]{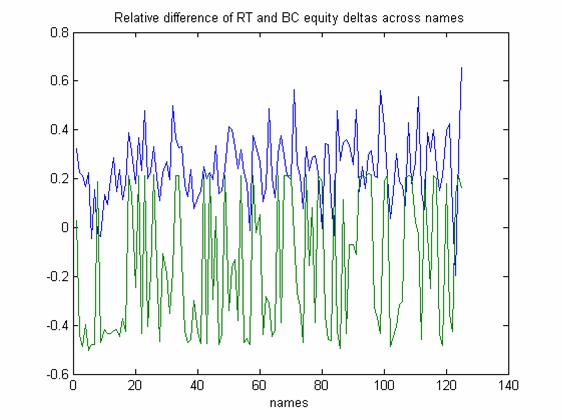}
\caption{Relative difference of equity delta obtained with 
the random thinning (RT) and base correlation (BC) method, for the 
``linear'' and ``base correlation'' choice of the prior. } 
\label{fig:EquityDelta}
\end{center}
\end{figure}

Going back to comparison of sensitivity parameters obtained with different 
RT schemes (see Tables ~\ref{Baxter}-\ref{Sprint}), 
the numbers shown there for three selected names 
do not really demonstrate that sensitivities 
obtained with the ``base correlation prior'' are considerably 
closer to base correlation sensitivities than those obtained 
with the linear prior. That this is indeed the case is illustrated 
in Fig.~\ref{fig:EquityDelta} where we show the relative difference
between the RT and BC equity delta across different names in the portfolio for
both the ``linear'' and ``base correlation'' choice for the 
prior\footnote{On average, RT sensitivities obtained with Prior II are in fact
somewhat closer to base correlation deltas than those obtained with Prior I, but 
this difference does not appear significant.}. 
What can be clearly seen though is that with the first prior our equity delta
is generally higher than the BC delta, while with ``base correlation'' prior 
the situation is reversed.  

Thus, our numerical experiment shows that the idea of having a sort of 
``calibration'' of TD matrices to base
correlation deltas does not really work, as the the subsequent rescaling
of TD matrices (needed to match single name and portfolio data) substantially 
alternates the prior matrices. While we are not aware of any compelling 
financial explanation of this behavior, 
perhaps a more interesting practical conclusion that can be drawn from 
our experiments with different priors is that single name deltas are not unique,
and depend on fine details of the model (in our case, the choice of the prior).
As will be further discussed in the concluding section, this situation is in fact
quite common, and occurs not only in top-down but in bottom-up models as well.

\section{From statics to dynamics: filtering approach}\label{SectionDynamics}

To get the single name {\it dynamics}, we need 
to update the TD-matrices dynamically based on observed information.
Updating of TD-matrices is done in two ways: by zeroing out rows and columns corresponding 
to ``observed'' (simulated) defaults, and by random perturbation (done in a particular way, see
below) of non-zero elements, to account for single name spread volatility.
Recall that if our TD-matrices are to be used for pricing today, they must be updated 
in a fair way: in other words, they must be martingales.

In this section, we consider a particular framework for modeling the filtration $ \mathcal{F}_t $.
Recall from Sect. \ref{SectionTD}.1, the filtration $ \mathcal{F}_t $ can be obtained by taking a combination of the natural top-down model filtration (i.e. default times and losses upon defaults,
but not defaulters' identities), the history of defaulters' identities (obtained by 
simulation, see below), and the filtration generated by observation of {\it 
``information processes''} (see below) for all names. These three components
of the filtration $ \mathcal{F}_t $ will be used below update the TD-matrices based on
portfolio losses, and the ``information process'' dynamics.


\subsection{Default history filtration}

In this section we consider updating our TD-matrices based on defaults in the portfolio,
i.e. adaptation of TD-matrices to the natural filtration of a top-down model. This is the 
most basic form of updating, and is necessary for simulation of defaulters' 
identities. We assume for now that there is no updating based on ``information 
processes'' (see the next section) and so in the notation of Sect.\ref{SectionTD}.1, the filtration $ \mathcal{F}_t $ is generated purely by the history of the portfolio-level default counting process, $(\mathcal{G}_t)_{t\ge0}$, and the identities of the defaulters, $(\mathcal{I}_t)_{t\ge0}$. 
In this section we replace the time 
dependence of the TD-matrices (representing the fact that we condition of $\mathcal{F}_t$) 
by a dependence on the number of defaults up to time $t$, $k \in \{ 0, 1,2, \ldots, N \}$. 

We assume that the initial portfolio contains no defaulted names.
Let us denote the calibrated conditional $TD$-matrices
as $ p_{ij}^{(m)}(k=0) $. Assume we simulate the portfolio dynamics using Monte Carlo, and the first default in the portfolio happens at time $ \tau^1$.
Conditional on the fact that a default has occurred, we independently simulate the identity 
of the defaulter with probabilities that depend on the relation between 
$ \tau^1 $ and the reference maturities $ T^{(m)} $. Let us first assume that
$ \tau^1 < T^{(1)} $. In this case, we simulate the identity $I_1\in \{ 1,2,\ldots, N \}$ of the first defaulter 
according to the conditional distribution $\{p_{i1}^{(1)}(k=0), i=1,\ldots,N\}$:
\beq
\label{condProb}
	P(I_1=i | \tau^1 < T^{(1)}) = p_{i1}^{(1)}(k=0).
\eeq
Note that we have some freedom here: instead of simulating from conditional probabilities
as determined by the TD-matrices at time 0, we may alternatively simulate from TD-matrices that have been dynamically updated as described in the next section to the time of actual default.
In the case when the first default happens in the interval $ [T^{(m-1)}, \, T^{(m)} ] $
with $ m > 1 $, we have:
\beq
\label{condProb2}
	P(I_1=i | T^{(m-1)} \le \tau^1 < T^m) = p_{i1}^{(m)}(k=0).
\eeq
Note that the method in (\ref{condProb}) (or (\ref{condProb2})) is similar to that
used previously by Duffie \cite{Duffie} (in the context of a bottom-up approach) to 
probabilistically pick a defaulter identity in a 
first-to-default basket where the simulated process is the 
aggregate portfolio default intensity. Unlike Duffie who employs a one-period setting, in 
our framework the conditional 
probabilities $ p_{i}^{(m)} $ are adapted to the model filtration in a dynamic way.

Conditional on the fact of ``observing" (for this particular Monte Carlo scenario) default
of name $ I_1 $, we now want to update our p-matrices $ p_{ij}^{(m)}(k=0) \rightarrow 
p_{ij}^{(m)}(k=1) $. In other words, we want to calculate conditional probabilities
\beq
\label{condProb3}
p_{ij}^{(m)} (k=1) = P \left[  \tau^j = \tau_i  | 
T^{(m-1)} \leq \tau_i \leq T^{(m)}, I_1, \tau^1 \right]\, , \; j = 2,\ldots,N, \, 
i = 1,\ldots,N.
\eeq   
We use the simplest possible ``model'' for these conditional probabilities:
\bea
\label{condProbModel}
p_{ij}^{(m)} (k=1) = \left\{ \begin{array}{lll} P \left[  \tau^j = \tau_i | T^{(m-1)} \leq 
\tau_i \leq T^{(m)}, \tau^1 \right] & \mbox{if $i \neq I_1 $} \\
0 & \mbox{if $ i = I_1$}.
\end{array}
\right.
\eea
In words, we assume that all rows with $ i \neq I_1 $ in 
the updated P-matrix do not depend on the identity of the first defaulter for all $ 
j \geq 2 $, while the row with $ i = I_1 $ is zeroed out. Note that our choice is
related to the 
fact that the natural filtration of the top-down model only contains default times but 
not identities of defaulters. Therefore, anything more complicated than 
(\ref{condProbModel}) would amount to some sort of a bottom-up, rather than top-down, 
approach.

What remains now is to come up with a method to calculate the conditional probabilities
in (\ref{condProbModel}). Again, note that we aim at a {\it simplest} possible 
dynamic model consistent with the portfolio-level dynamics of the top-down model and its natural 
filtration. To this end, we note that the new updated probabilities should 
still satisfy the first of constraints (\ref{marginals}), where the new right hand side
can now be calculated for all $ j \geq 2 $ using the top-down model. Respectively, we obtain a 
simplest model for (\ref{condProbModel}) by a uniform rescaling of all probabilities 
$ p_{ij}^{(m)}(k=0) $ for all $ j \geq 2 $ and $ i \neq I_1 $ so that the first constraint
in (\ref{condConstraints}) is re-enforced:
\beq
\label{rescaling}
p_{ij}^{(m)} (k=1) = \frac{p_{ij}^{(m)} (k=0)}{
1-p_{I_1j}^{(m)}(k=0)} \; , \; \; i \neq I_1, \; j = 2, \ldots, N.
\eeq

Note that after this rescaling is done, we get 
{\it different} marginal forward default probabilities     
$ P \left[ \left. T^{(m-1)} \leq \tau_i  \leq T^{(m)} 
 \right| \mathcal{F}_t \right] $ for surviving names, which are 
obtained if we sum the product of the new conditional probabilities $ p_{ij}^{(m)}(k=1) $ with the newly calculated tail probabilities over $ j $. 
Extending the above analysis to the second, third, etc. defaults, we
have the following scheme for simulation of defaulters' identities and updating the conditional p-matrices. For each 
sequential default, we simulate the defaulter's identity from the current p-matrices, 
as in (\ref{condProb}), and then zero out the corresponding column and row
in the current p-matrices. We then rescale the current
p-matrices so that the column-constraints are satisfied. The new re-scaled p-matrices
are used in conjunction with the tail probabilities $w_k^{(m)}(j)$ to calculate the new forward marginal default probabilities for surviving names.

Qualitatively, the impact of defaults on default intensities of surviving names
can be described as an interplay between three effects. To explain them, consider 
a particular scenario where we observed $n$-th default at time $ t $. Assuming the
identity of this defaulter is $ I_t $, we zero out the $I_t $-th row in all
forward matrices, and rescale all columns of $ p_{ij} $ for all $ j > n $ to re-establish the column
constraints. This has an effect of bumping all probabilities $ p_{ij} $ for any fixed
$ j > n $. However, we have also to take into account the fact that when a default 
occurs, the ``next-to-default'' column in the $p$-matrices moves one step to the 
right. Therefore, the impact of this default on short-term spreads of surviving names
will be determined by a combined effect of three factors: 1) we move one column to the 
right in the $p-$ (or $ P-$) matrices, 2) we rescale this column to re-enforce the 
column constraint, and 3) we multiply the result by the new portfolio-level intensity
$ \lambda_{n} $ corresponding to $n$ defaults in the portfolio, instead of a previous
multiplication by $\lambda_{n-1} $. As $ \lambda_n $ is generally an increasing 
function of $ n $, the net effect will be largely driven by the first two factors.
This implies, in particular, that for the simple model with a ``uniformization bound'' for names with low spread we will have a relation $ p_{i,j+1} > p_{ij} $, so the net 
effect will be to increase the default intensity. For names with high initial spreads,
we have an inverse relation $ p_{i,j+1} < p_{ij} $, and therefore the resulting change
of their default intensity can be of any sign, depending on the relative strength of 
two effects. This is largely consistent with what is observed in the markets: upon
defaults, low-spread names (especially those in the same industry/sector) are expected
to go up due to informational contagion, while high spread names do not necessarily 
widen - they may already be wide enough to start with.

\begin{figure}[ht]
\begin{center}
\includegraphics[width=60mm,height=50mm]{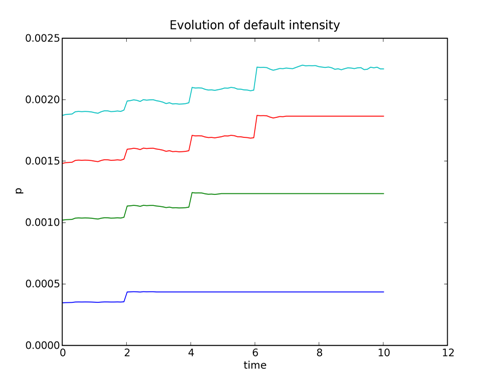}
\includegraphics[width=60mm,height=50mm]{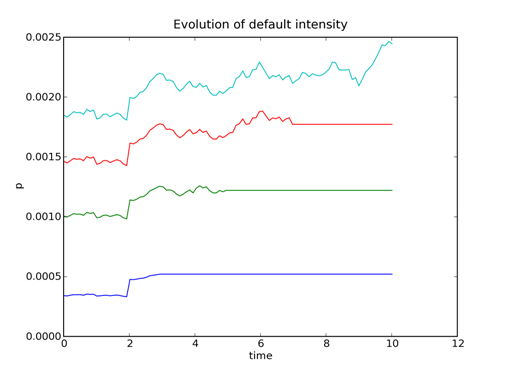}
\caption{Default updating of conditional TD-matrices: a row corresponding 
to a low spread names for matrices $ p_{ij}^{(m)} $ for all maturity
intervals $ m = 1, 2,3,4 $. (Note that each line flattens out at the end of 
its maturity interval. The lowest and top-most lines correspond to $ m = 1 $
and $ m = 4 $, respectively.)
The two graphs correspond to regimes of low and 
high volatility of background ``information'' process. One sees that adding volatility does 
not completely wash out jumps resulting from the default updating.
} 
\label{fig:Default_update_low_spread}
\end{center}
\end{figure}

\begin{figure}[ht]
\begin{center}
\includegraphics[width=60mm,height=50mm]{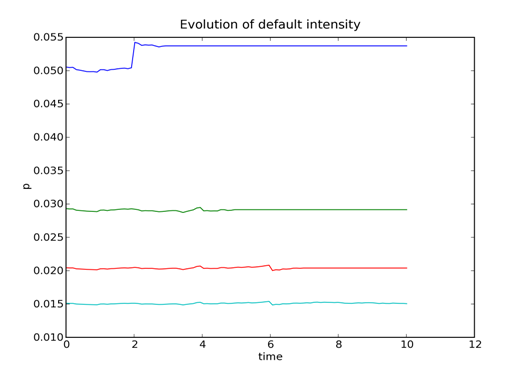}
\includegraphics[width=60mm,height=50mm]{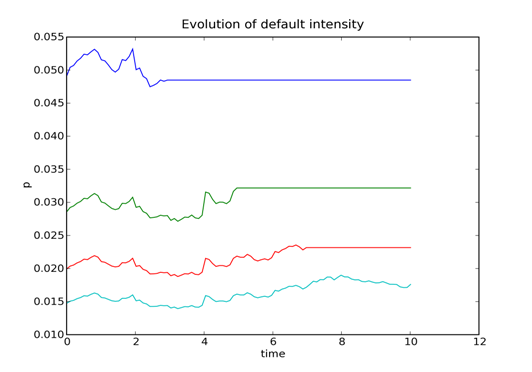}
\caption{Default updating of conditional TD-matrices: a row corresponding 
to a high spread names. (Note that ordering of lines is reversed in comparison to 
Fig.~\ref{fig:Default_update_low_spread}: the top-most line corresponds to $ m = 1 $,
and the line at the bottom corresponds to $ m = 4 $.) 
The two graphs correspond to regimes of low and 
high volatility of background ``information'' process. One sees that the only clearly seen
jump occurring for matrix $ p^{(1)}_{ij} $ is washed out when volatility is added.} 
\label{fig:Default_update_high_spread}
\end{center}
\end{figure}

Examples of default updating of low spread and high spread names for a scenario
with three defaults happening at 2, 4 and 6 years are displayed in 
Figs.~\ref{fig:Default_update_low_spread} and \ref{fig:Default_update_high_spread}
where we show updating of a single row
in conditional TD-matrices upon portfolio defaults for 
low- and high-spread names, respectively\footnote{In addition to 
portfolio defaults, matrices are updated to a background ``information'' process in
a way explained below in Sect. \ref{SectionDynamics}.2. Two figures for each case correspond to regimes 
of low and high volatility of information process.}. The resulting jumps in single name
intensities (as expected, more pronounced for a low spread name) manifest  
the credit contagion at the single name level.

\subsection{Spread dynamics}


Applying the default updating procedure above leads to matrices $p_{ij}^{(m)}$ that are piecewise constant between defaults. From Sect.\ref{SectionTD}, this implies that the forward default probabilities of the single names are perfectly correlated with the index, which is unrealistic. In order to remedy this, we must condition on additional information; in addition to observing defaults and the identities of defaulting names, we now assume that for each name $ i $, we observe 
an {\it ``information process''} $X_i$ 
driven by a Brownian motion\footnote{ 
Our ``information process'' is similar to that introduced by Brody, Hughston and Macrina (BHM)
\cite{BHM}, but we are not forced to use a Brownian bridge in our modeling
framework, and can instead proceed with the usual Brownian motion.}.
The filtration  $(\mathcal{H}_t)_{t\ge0}$ introduced in Sect. \ref{SectionTD}.1 is assumed to be 
generated by the joint history of the information processes  $ \{ X_i \}$ for all names. 
We note that the processes 
$ \{ X_i \} $ may be correlated with each other, and for each name $ i $, its
process $ X_i $ can be correlated with market observables such as the equity price.
While many functional forms for $X_i $ are possible (e.g. $ X_i$ could be a GBM or OU process),
we restrict ourselves below to a simplest case of a Brownian diffusion with an unknown random drift.
  
To model stochasticity in the TD-matrices, we assume that 
information about the order of defaulting names 
(first to default, etc) is encoded in the drifts of 
the $X_i$'s, while the accuracy of our information about the $i$-th name is described 
by the volatility of the $ X_i$. For simplicity, we concentrate on 
modeling stochasticity 
of the next-to-default intensities (i.e. the nearest ``active'' columns 
$  p_i^m(t) \equiv p_{ik}^{(m)}(t) $ 
(with $ i = 1, \ldots, N $) of the conditional TD-matrices in a 
scenario where we have $ (k-1)$ previous defaults).  
 
We assume that for a given Monte Carlo scenario,
some names have already defaulted by a previous 
time $ s < t $, and that $ p_i^m(s) $ was already adapted to
filtration $ \mathcal{F}_{s} $.   


Let $ \bf{X}_t $ be an $n$-dimensional It{\^o} process described by the following 
 stochastic differential equation (SDE): 
\beq
\label{obsProcess}
d {\bf X}_t = {\bf{\Theta}}_t dt +  \Gamma  d \bf{W}_t , 
\eeq
where the volatility matrix $ \Gamma $, with elements $ \Gamma_{ij} = \sigma_{ij} $,
satisfies the constraints
\beq
\label{Gamma}
\left( \Gamma \Gamma' \right)_{ij} = \Sigma_{ij} = \rho_{ij} \sigma_i \sigma_j, 
\eeq
(here $ \Sigma $ stands for the covariance matrix), and 
the $n$-dimensional drift $ {\bf{\Theta}}_t $ follows a 
$N$-state continuous time Markov chain over 
states  $ \{ {\bf{\Theta}}_1, {\bf{\Theta}}_2, \ldots, {\bf{\Theta}}_N \} $ with 
generator matrix $ \Lambda $. In our case, we set $ n = N $, i.e. the dimension
of the information process is equal to the size of the portfolio. We will, however, keep 
using different symbols $ N $ and $n $ in the formulas below in order to clarify 
whether we operate on hidden $ (N)$ or observable $ (n) $ degrees of freedom.  
The unobservable vector-valued drift
$ \bf{\Theta}_t $ with values  
\bea
\label{muVec}
{\bf{\Theta}}_i = \left( \begin{array}{clcr}
\Theta_{i1} \\
\Theta_{i2} \\
\vdots \\
\Theta_{in}
\end{array} \right) =
\left( \begin{array}{clcr}
\delta_{i1} \\
\delta_{i2} \\
\vdots \\ 
\delta_{in} 
\end{array} \right)\; , \; \; i = 1,\ldots, N,
\eea
describes the identity of the next defaulter: if name 1 is to default next, then
$ {\bf{\Theta}} = {\bf{\Theta}}_1 = (1,0,\ldots,0)' $, if name 2 is the next defaulter, then
$ {\bf{\Theta}} = {\bf{\Theta}}_2 = (0,1,\ldots, 0)' $, etc. 

We assume 
that the market filtration $ \mathcal{F}_t $ 
is generated by continuous observation of the process
$ \bf{X} $. We are therefore faced with the problem of inference on a hidden state
$ {\bf \Theta}_t $ given the observation of $ \bf{X} $. In other words, we want to calculate 
the posterior probability
\beq
\label{posterior}
\pi_i(t) = P \left[ \bf{\Theta}_t = \bf{\Theta}_i | \mathcal{F}_t \right],
\eeq
given the prior probabilities 
\beq
\label{priorProb}
\pi_i(0) = \hat{\pi}_i.
\eeq

\subsubsection{Continuous time formulation}

As shown by Liptser and Shiryaev \cite{LS}, the posteriors (\ref{posterior})
satisfy the following system of stochastic differential equations:
\beq
\label{LSeq}
d \pi_i(t) = \sum_{j=1}^{N}  \pi_j(t) \lambda_{ji} dt + 
\pi_i(t) \hat{\Gamma}_{i} d \tilde{\bf{W}_t}.   
\eeq
Here $ \hat{\Gamma}_{i} = \left( \hat{\Gamma}_{i1}, \ldots,  \hat{\Gamma}_{in} \right)$  with elements 
\beq
\label{volBelief}
\hat{\Gamma}_{ij} = \frac{\Theta_{ij} - \bar{\bf{\Theta}}_j}{\sqrt{\sum_{m=1}^{n}
\sigma_{jm}^2}} =  \frac{\delta_{ij} - \bar{\bf{\Theta}}_j}{ \sigma_j} , 
\eeq
is the matrix of volatilities of beliefs,
\bea
\bar{\bf{\Theta}} =  \left( \begin{array}{clcr}
\bar{\Theta}_1 \\
\bar{\Theta}_2 \\
\vdots \\
\bar{\Theta}_n \\
\end{array} \right) = 
\sum_{i=1} {\bf{\Theta}}_i \pi_i(t) = 
\sum_{i=1} \pi_i(t)  \left( \begin{array}{clcr}
\Theta_{i1} \\
\Theta_{i2} \\
\vdots \\
\Theta_{in}
\end{array} \right) = 
 \left( \begin{array}{clcr}
\pi_1 \\
\pi_2 \\
\vdots \\
\pi_n
\end{array} \right),
\eea
is the expected drift, and 
$ \tilde{\bf{W}}_t = \left( \tilde{W}_t^1, \ldots, \tilde{W}_t^n \right)' $ 
is a $ \mathcal{F}_t $-adapted $n$-dimensional Brownian motion (``innovation process'')
with components 
\beq
\label{Wi}
d \tilde{W}_t^j = \frac{1}{\sqrt{\sum_{k=1}^n \sigma_{jk}^2}} \left( 
d X_t^j - \bar{\Theta}_j dt \right) \; , \; \; j = 1, \ldots, n,
\eeq 
that is
assumed to be observable by investors. Its covariance is calculated as follows:
\beq
\label{covInno}
\la d \tilde{W}_t^j d \tilde{W}_t^k \ra = \frac{\sum_{n} \sigma_{jn} \sigma_{kn}}{
\sqrt{ \sum_{n} \sigma_{jn}^2 \sum_{m} \sigma_{km}^2 }} \, dt \equiv 
 \frac{ \left( \Gamma \Gamma' \right)_{jk}}{ \sigma_j \sigma_k } \, dt = \rho_{jk} dt, 
\eeq
where we used (\ref{Gamma}). The instantaneous 
lognormal volatility $ \hat{\sigma}_i $ of $ \pi_i $ is given by:
\beq
\label{instVol}
\hat{\sigma}_i^2 = \sum_{j,k} \hat{\Gamma}_{ij} \hat{\Gamma}_{ij} \frac{
\la d \tilde{W}_t^j d \tilde{W}_t^k \ra }{dt}  = 
\sum_{j,k}  \frac{ \delta_{ij} - \pi_j }{ \sigma_{j}} 
 \frac{ \delta_{ik} - \pi_k }{ \sigma_{k}} \rho_{jk}.    
\eeq 
In our setting, we take $ \Lambda = 0 $, i.e. we consider 
the identity of the next defaulter to be 
a random variable rather than a random process. Eq. (\ref{LSeq}) in this case reduces
to 
\beq
\label{LSeq1}
d \pi_i(t) =  \pi_i(t) \hat{\Gamma}_{i} d \tilde{\bf{W}_t}.   
\eeq
A few comments on the structure of Eqs.(\ref{LSeq1}) and (\ref{LSeq}) 
are in order here. First, we observe
that the posterior probabilities $ \pi_i $ specified by (\ref{LSeq1}) are martingales. 
Second, both Eqs.(\ref{LSeq}) and (\ref{LSeq1}) ensure 
conservation of probability $ \sum_{i} d \pi_i(t) = 0 $. Third, we note that the diffusion
term in both (\ref{LSeq}) and (\ref{LSeq1}) vanishes in both limits $ \pi_i \rightarrow 0 $ and  $ \pi_i \rightarrow 1 $: the first result holds as the diffusion term is 
proportional to  $ \pi_i $, and the second results follows by the structure of 
(\ref{instVol})\footnote{When $ \pi_i \rightarrow 1 $, we have $ \pi_j \rightarrow 0 $ for 
all $ j \neq i $, and thus the factor $ (\delta_{ij}- \pi_j) $ vanishes for all $ j $.} . 
Lastly, note that in the limit where  
$ \sigma_i \rightarrow \infty $ for all $ i $, Eq. (\ref{LSeq}) reduces to the forward 
Kolmogorov equation as it should, since in this case, observation of the ``information process'' is useless for inference of the hidden Markov chain.
  
\subsubsection{Discrete time formulation}

In practice, simulation using Eq. (\ref{LSeq}) or (\ref{LSeq1}) can be somewhat tricky as 
discretization may lead to spurious negative probabilities, non-conservation of probability, etc,
similar (but more involved) to problems arising in discretization of e.g. a CIR process.
In this view, it may still be preferable to use a discrete-time formulation based on 
Bayes' theorem for simulation, while retaining the SDE formulation for calibration
(see below on this).

Let  $ D_t $ be the data that becomes
available at time $ t $; in our case $ D_t = \{ X_t^i \} $).     
Therefore, we use Bayes' theorem 
\beq
\label{Bayes}
p( I_{next} = i | D_t, \mathcal{F}_s) = p( I_{next} = i | \mathcal{F}_s) \frac{ 
P(D_t|I_{next} = i, \mathcal{F}_s)}{\sum_{i} p(I_{next} = i | 
\mathcal{F}_s) P(D_t|I_{next} = i, \mathcal{F}_s)}, 
\eeq
to adapt our probabilities to filtration generated by information 
processes $ \{ X_t^i \} $. We thus get the updating rule:
\beq
\label{updating}
p_i^m(t+ \Delta t) = p_i^m(t) \frac{P \left[ \Delta {\bf X}_t | {\bf \Theta}_t = 
{\bf \Theta}_i \right] }{
\sum_{i} P \left[ \Delta {\bf X}_t | {\bf \Theta}_t = {\bf \Theta}_i \right] P \left[ 
{\bf \Theta}_t = {\bf \Theta}_i \right]},
\eeq
where
\beq
\label{likeli}
 P \left[ \Delta {\bf X}_t | {\bf \Theta}_t = {\bf \Theta}_i \right] \sim 
 \exp \left( - \frac{1}{2 \Delta t} 
\left( \Delta {\bf X}_t - {\bf \Theta}_i \Delta t \right)^{T} \Sigma^{-1}
\left( \Delta {\bf X}_t - {\bf \Theta}_i \Delta t \right) \right).
\eeq
This can be generated by simulating  
the defaulter identity $ k $ and a standard normal random variable $ Z_{t}^j $: 
\beq
\label{decomp}
\Delta {\bf X}_t = \Delta t {\bf \Theta}_k + \sqrt{\Delta t} {\bf W}.
\eeq
Simplifying, we obtain
\beq
\label{updating2}
p_i^m(t+ \Delta t) = \frac{ p_i^m(t) L_i \left( \Delta {\bf X}_t \right)}{
\sum_{i} p_i^m(t) L_i \left( \Delta {\bf X}_t \right)},
\eeq
where the likelihood $L_i$ is given by
\beq
\label{Li}
 L_i \left( \Delta {\bf X}_t \right) = \exp \left[ \left( \Sigma^{-1} \Delta {\bf X}_t 
\right)_i - \frac{\Delta t}{2} \Sigma^{-1}_{ii} \right].
\eeq

\section{Numerical algorithm}\label{SectionNumerical}
%

Start by setting $k=1$, where the index $ k $ enumerates the number of the next default. Then for $k\ge 1$:
\begin{enumerate}
\item Default Simulation ($\tau^k$): Given $k\ge 1$, simulate $\tau^k$ according to the top-down model, and find the index $m$ of the forward interval such that $\tau^k \in [T^{(m-1)},T^{(m)}]$.
\item Identity Simulation ($I_k$): Simulate the identity of the $k$th defaulter by sampling from the $k$-th column of the p-matrix for the $m$-th forward interval.
\item Information Updating (on $[\tau^{k-1},\tau^k]$, optional): Use information process filtering to update the p-matrices to time $\tau^k$. The Bayesian update ensures the dynamics of the p-matrices are consistent with the identity simulation step above.
\item Default Updating (at $\tau^k$): This step ensures that future identities are simulated consistently with the identity simulation step above (i.e. no repeat defaulters).
\begin{enumerate}
	\item Zero out the $ I_k $-th row of all p-matrices.
	\item Zero out the $k$-th column of all p-matrices.
	\item Re-enforce column constraints for the p-matrices.
\end{enumerate}
\item Repeat: If $ \tau^k $ is less than the final maturity, 
increment $ k \rightarrow k + 1 $ and go to Step 1.
\end{enumerate}

Note that as soon as some $\tau^k > T^m$, the $m$-th p-matrix can be dropped from any 
future updating steps, as the simulation has progressed beyond the point at which this 
matrix applies.

\section{Marking parameters of the information updating}\label{SectionMarking}

As discussed above, volatility and correlation parameters of the information
processes impact the volatilities and correlation of the resulting single name spreads. 
Therefore, one can attempt to mark those parameters in such
a way that the model-implied behavior of single name spreads will be roughly consistent
with the empirical behavior. In this section we present a semi-quantitative analysis 
toward this goal. Note that there are a few reasons why we will not discuss
an accurate calibration. First, empirical variances and especially covariances   
are subject to large measurement errors, and ideally should be ``noise-undressed'' before 
the use. Second, the relation between vols and correlations of the driving Brownian 
motions and vols and correlation of observable spreads is highly non-linear and hard to 
work with unless a number of approximations are made.

In principle, the task of approximate calibration/marking of correlation parameters can
be done using two alternative approaches. The first one is to work with probabilities, 
i.e. conditional forward TD matrices and tail probabilities coming from the top-down model.
The second way is to use spot default intensities as proxies for finite maturity spreads,
and work with them. We choose the second route as it appears simpler technically.

Consider again formulas (\ref{Z2}) and (\ref{ansatz}), which we repeat here for 
convenience :
\beq
\label{Z2copy}
\lambda_t^i = Z_t^i \lambda_t^p \; , \; \; \
Z_t^i = p_{i,N_t}^{(1)} 1_{(t \leq T_1)} + p_{i,N_t}^{(2)} 1_{(T_1 < t \leq T_2)}
+ \ldots 
\eeq
Assume that we want to calibrate volatility structure of Brownian motion in the no-default environment for $ t < 3Y $. We can then identify 
the thinning parameters $ Z_t^i $ with the first column $ \pi_i \equiv p_{i1}^m$ of 
the first ($ m = 1$) conditional $p$-matrix.
From (\ref{Z2copy}) we obtain (omitting the time index and 
denoting $ \lambda_t^p \equiv \lambda $ to ease the notation)
\beq
\label{dlambda}
\frac{d \lambda_i}{\lambda_i} = 
\frac{\pi_i d \lambda + \lambda d \pi_i}{ \pi_i \lambda } = \frac{d \lambda}{\lambda} +
\frac{d \pi_i}{\pi_i}, 
\eeq
which yields
\beq
\label{covLambdas}
\la \frac{d \lambda_i}{ \lambda_i} \, \frac{d \lambda_k}{\lambda_k} \ra = 
\la \left( \frac{ d \lambda}{\lambda} \right)^2 \ra + \la \frac{d \lambda}{\lambda} \, 
\frac{d \pi_k}{\pi_k} \ra + \la \frac{ d \pi_i}{ \pi_i} \, \frac{d \lambda}{\lambda}
\ra + \la \frac{d \pi_i}{ \pi_i} \, \frac{ d \pi_k}{\pi_k} \ra .
\eeq 
As we assume independence between the $ Y$-process and Brownian motion 
$ \tilde{{\bf W}}_t $, the cross terms in this formula vanish, and we obtain
\beq
\label{covSprdConstr}
C_{ij}^{\pi} dt \equiv \la \frac{d \pi_i}{ \pi_i} \, \frac{ d \pi_j}{\pi_j} \ra  = 
\la \frac{d \lambda_i}{ \lambda_i} \, \frac{d \lambda_j}{\lambda_j} \ra
- \sigma_Y^2 dt. 
\eeq
Note that this matrix is orthogonal to the vector $ \pi $:
\beq
\label{orthog}
\sum_{j} C_{ij}^{\pi} \pi_j dt = \la \frac{d \pi_i}{ \pi_i} \sum_{j} d \pi_i \ra  dt 
= 0, 
\eeq
because $ \sum_{j} d \pi_j = 0 $ according to Eq. (\ref{LSeq1}).
    
We use (\ref{covSprdConstr}) as a 
constraint on the covariance of posterior probabilities 
$ \pi_i(t) $ by approximating the first term in the right hand side by the empirical 
covariance matrix of spread returns (times $ dt $)\footnote{Note that by doing this,
we tacitly assume time stationarity of a process driving credit spreads.}. On 
the other hand, we can 
calculate the covariance matrix of $ d \pi_i / \pi_i $ using Eq. (\ref{LSeq1}):
\beq
\label{covPi}
\la \frac{d \pi_i}{\pi_i} \, \frac{d \pi_k}{\pi_k} \ra = \sum_{j,l} \hat{\Gamma}_{ij}
\hat{\Gamma}_{kl} \la d \tilde{W}_t^j \, d \tilde{W}_t^l \ra = \left( \hat{\Gamma} 
\rho \hat{\Gamma}' \right)_{ik} dt \equiv \left( \gamma \mathcal{R} \gamma' 
\right)_{ik} dt,
\eeq
where $ \gamma_{ij} = \delta_{ij} - \pi_j $ and $ \mathcal{R}_{ij} = 
\frac{\rho_{ij}}{\sigma_i \sigma_j} $. Note that matrix $ \gamma $ is degenerate, 
therefore we cannot directly invert Eq. (\ref{covPi}) to find matrix $ 
\mathcal{R} $ from $  C^{\pi} $. However, a solution to this equation can be 
easily found as long as the orthogonality relation $ C \pi = 0 $ holds. Indeed,
in this case we find
\beq
\label{selfSol}
\gamma C \gamma' = \left( 1 - 1 \cdot \pi' \right) C \left( 1 - \pi \cdot 1' \right) 
= C,
\eeq
which shows that $ \mathcal{R} = C $ is a solution of (\ref{covPi}) as 
long as $ C \pi = 0 $. 
Note that in theory, matrix $ C $ is indeed orthogonal to $ \pi $ as shown 
in (\ref{orthog}). However,
in our treatment we approximate matrix $ C $ using observable spreads, see again 
(\ref{covSprdConstr}). In doing so, the orthogonality property is generally lost,
and thus such an estimated matrix $ C $ does not solve (\ref{covPi})
anymore. 

This observation prompts a way of solving Eq. (\ref{covPi}) 
{\it approximately} (in the least square sense) rather than exactly. Namely, for 
a given empirical matrix $ C $, we find a matrix $ \hat{C} $ such that $ \hat{C} $
is as close as possible to $ C $ (e.g. in the sense of Frobenius norm), while on the 
other hand satisfying the orthogonality constraint $ \hat{C} \pi = 0 $.
Once such a matrix $ \hat{C} $ is found, the approximate solution of (\ref{covPi}) is
given by $ \mathcal{R} = \hat{C} $.

A candidate solution satisfying the orthogonality relation can be 
presented\footnote{We thank Yury Volvovskiy for a discussion of Eq.(\ref{covPi}) and
suggesting the candidate solution (\ref{candidate1}).}:
\beq
\label{candidate1}
\hat{C}_{ij}^{(1)} = C_{ij} - \frac{ \left(C \pi \right)_i \left( \pi' \cdot C
\right)_j}{ \pi' C \pi},
\eeq
however it does not guarantee that matrix $ \hat{C} $ defined in this way is the 
closest possible to $ C$. The matrix closest to $ C $ is instead given by the 
usual projector
\beq
\label{candidate2}
\hat{C}_{ij}^{(2)} = C_{ij} - \frac{\pi_j}{ || \pi ||^2} \sum_{k} C_{ik} \pi_k. 
\eeq
However, such a matrix $ \hat{C} $ cannot be the right solution as it is non-symmetric.
We have therefore set up a recursive algorithm where we alternate between the first 
step that calculates 
the orthogonal projection (\ref{candidate2}), and the second step 
where the resulting matrix is made symmetric and positive definite by truncating
(zeroing out) any possible negative eigenvalues. As the second step can be 
viewed as a projection on a convex set of positive-definite symmetric matrices,
our procedure amounts to alternating projections on two convex sets, which is known 
to converge. 

In practice, with our particular dataset that we used to test our method (see below),
we found that doing 10 to 20 iteration of such procedure provides a 
matrix that is approximately orthogonal to $ \pi $ with a good accuracy, while 
being symmetric 
and positive definite. The Frobenius norm of the difference $ || \hat{C} - C ||^2 = 
\sum_{ij} \left( \hat{C}_{ij} - C_{ij} \right)^2 $
calculated with this method was found to be about 3 times smaller than the norm of 
the difference calculated according to (\ref{candidate1}).

To check accuracy of these approximations and to provide a numerical example, we have 
analyzed one year of daily 
tranche and single name data for CDX IG8 ending at 03/03/08. We present our results 
for averages across all names in the portfolio, rather than individual names. 
Using (\ref{covSprdConstr}) for all names and assuming $ \sigma_Y = 0.35 $, we 
find that the portfolio-average
volatility of $ \{ \pi_i \}  $ is 1.71, with a minimum of 1.33, maximum of 7.68,
and the standard deviation (std) of 0.6. The portfolio-average correlation between
different $ \pi_i$'s is 0.58, with std of 0.13\footnote{Note that for 
the lack of relevant risk-neutral data, we use here historical values for single name vols and 
correlations, while the value $ \sigma_Y = 0.35 $ is chosen to be approximately equal to 
the implied index volatility as obtained from short-dated (6M and 1Y) index options. We are 
interested, of course, in estimations relevant for the risk-neutral world.}.
We then use the procedure described above to calibrate the parameters of the Brownian
motions. This produces the average volatility of 0.68, with the minimum and maximum
of 0.13 and 1.93, respectively, and std 0.08. The mean correlation of Brownian motions
is 0.48 and std is 0.28. We then simulate the information updating scheme 
using these parameters
of the Brownian motion, and compute output volatility and correlations of 
posterior probabilities $ \{ \pi_i \} $. 

Results are displayed in Table ~\ref{OutPi}. 
\begin{table}
\vspace{5mm}
\begin{tabular}{|c|c|c|c|r|}
\hline
numRuns & meanVol  & stdVol & meanCor & stdCor    \\
\hline
500 & $ 1.63 \pm 0.10  $ & 
$ 1.08 \pm 0.02 $ & $ 0.63 \pm 0.02 $ &  $ 0.34 \pm 0.01 $  \\
\hline
1000 & $ 1.64 \pm 0.08 $ & 
$ 1.09 \pm 0.01 $ & $ 0.62 \pm 0.01 $ &  $ 0.34 \pm 0.01 $ \\
\hline
\end{tabular}
\label{OutPi}
\caption{Monte Carlo output volatilities and correlations of posterior probabilities
$ \{ \pi_i \} $.}
\end{table}
One sees that the agreement between the 
input and output parameters is reasonable though not exact. Factors 
that contribute to the mismatch are 
noise in the data, possible non-stationarity of credit spreads dynamics, and potential 
loss of accuracy in proxying the covariance matrix of $ \pi_i$'s by the empirical covariance
matrix. Furthermore, we found it necessary to truncate small eigenvalues in the resulting
covariance matrix of Brownian motions calculated as described above\footnote{
Here by truncation we mean a SVD-type truncation, where 
we zero out small eigenvalues in the covariance matrix, and simultaneously zero
out corresponding inverse eigenvalues in the inverse matrix appearing in 
(\ref{Li}). The number of 
eigenvalues to keep was found empirically to be around 40 by looking for the best match
of input parameters by the Monte Carlo simulation.}. The need for this procedure
arises because when an unadjusted covariance matrix obtained as described above is used for 
simulation, in a substantial fraction of Monte Carlo scenarios we find the 
phenomenon of a
``super-fast learning'', where the identity of the next defaulter is 
learned too fast, so that
spread volatilities drop to zero after a short (1-2 years) 
initial period of high volatility, which is clearly not 
a desirable behavior. The origin of this behavior can be traced back to Eq.(\ref{Li}) defining
the likelihood function: when small eigenvalues are present in the covariance matrix, 
the likelihood becomes large, and the hidden state is learned almost with certainty. A 
truncation of small eigenvalues of the covariance matrix makes the process
more stationary and thus fixes this 
problem, albeit perhaps 
not in a way most attractive from a theoretical standpoint\footnote{We note that a 
similar 
phenomenon of a ``super-fast learning'' can also occur in the BHM model
\cite{BHM}, which is not surprising given the fact that both their and our 
approaches use the 
idea of learning a  fixed unknown {\it state}, rather than a dynamic {\it process}.}.
We would like to note, however, that adding actual 
defaults to sample paths is able to mitigate the effect
of super-fast learning as a name will likely default before filtering will show 
this name as a sure next defaulter. Therefore, we believe that in practice, 
incorporating defaults together with a truncation of eigenvalues provides a satisfactory
solution to the problem of possible super-fast learning. 
A few more related comments will be made in the next
section. 

\begin{figure}[ht]
\begin{center}
\includegraphics[width=60mm,height=50mm]{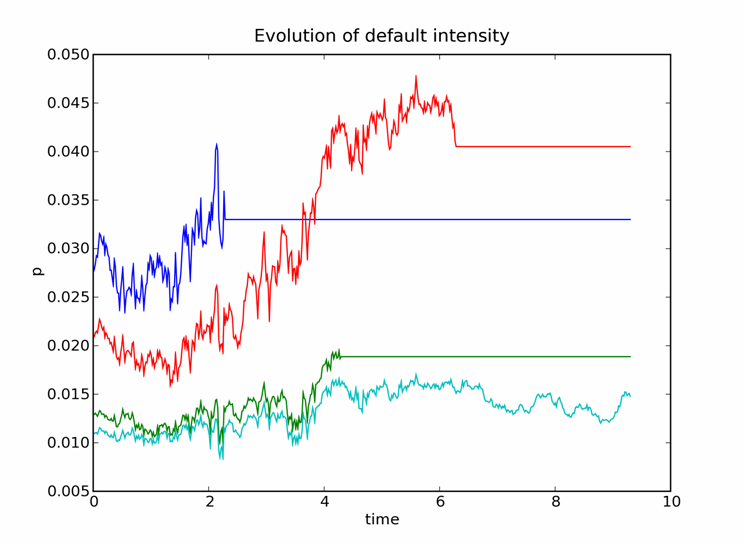}
\includegraphics[width=60mm,height=50mm]{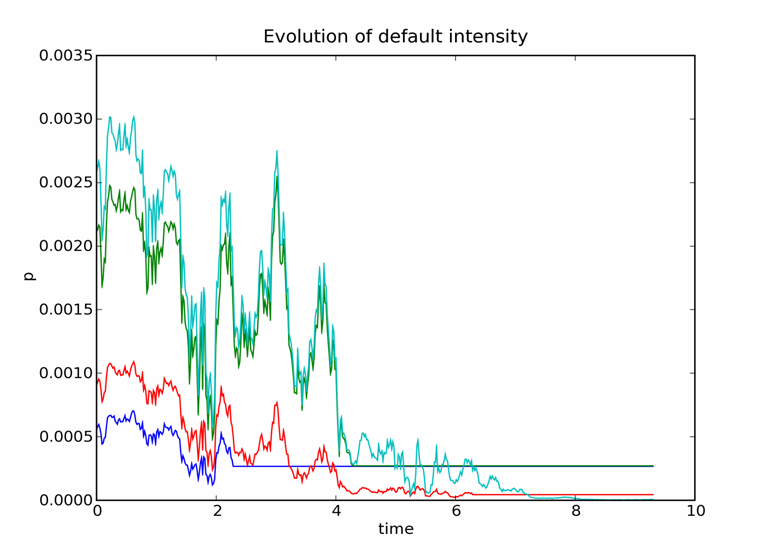}
\caption{Sample paths for simulated p-matrices with calibrated parameters
for high- and low-spread names (the left and right graphs, respectively). These
sample paths produce no portfolio defaults. Shown are the next-to-default elements
of corresponding rows of matrices $ p_{ij}^{(1)} $ to $ p_{ij}^{(4)} $.
} 
\label{fig:SamplePath}
\end{center}
\end{figure}

We show in Fig.~\ref{fig:SamplePath} sample paths of the short-term default intensity
(elements of the next-to-default columns in the conditional TD-matrices) for 
high- and low-spread names.

\section{Summary and discussion}\label{SectionSummary}

In summary, we have presented a practically-oriented random thinning (RT) framework,
with an attempt geared toward flexibility, accuracy and numerical efficiency.
In particular, our parametrization of TD-matrices is tailored to a very fast (as 
compared to more traditional inner point methods that would be required with a 
different parametrization) iterative scaling (IS) method. Furthermore, we have attempted 
to extend this approach to a dynamic single name setting by developing a scheme for the 
default and ``information process'' updating, and have presented a numerical 
scheme for marking model parameters. 

Our proposed dynamic scheme is not free of drawbacks. In particular, we are only able to
match CDS spread volatilities and correlations in the portfolio-averaged sense but not 
on the name-by-name basis. We believe that this still can be considered (modest) progress
in the right direction, given the fact that most, if not all, bottom-up models, either struggle
in calibrating to these data, or give it up altogether. Better control of  
volatilities and correlations of spreads in a basket is likely to be important for derivatives
that are sensitive to both spread and default dynamics of a credit portfolio.

Another drawback of our approach is a potential ``super-fast learning'' (see the discussion
at the end of last section). The possibility of a super-fast learning stems from a combination
of high volatilities of credit spreads (i.e. the current market environment), and 
the fact
that we learn a fixed, albeit unknown, hidden state, i.e. a random {\it variable}, 
not a random {\it process}\footnote{It appears that in the current setting of learning 
a hidden stage, a multi-dimensional distribution
with fatter tails than a Gaussian could help in preventing a super-fast learning.
We have experimented with a Student's t-distribution, but found no substantial 
improvement over the Gaussian case.}. We have found a 
way to semi-empirically tackle this issue by truncating
small eigenvalues of the Brownian covariance matrix, however a more principled approach
would be clearly desirable. A possibility of changing the scheme so that 
a hidden {\it process},
rather than a hidden {\it state} would be learned, in a way that respects 
no-arbitrage martingale
constraints, is currently an open question.

In conclusion, we would like to discuss a few further issues that are related to the 
model presented here. 

\subsection{Valuing bespoke tranches}

The problem of pricing and risk management of bespoke tranches off liquid index 
tranches is of acute interests for practitioners. The current industry-standard
methods typically involve some form of mapping base correlations for index 
tranches onto correlation numbers for the bespoke portfolio. As 
the index and bespoke portfolios typically have different overall risk levels,
the same values of strikes for the index 
and bespoke portfolio  have different meanings, which precludes the straightforward
use of base correlations to price bespoke tranches. Instead, practitioners 
rely on mapping the strikes of the two portfolio onto each other using some relative
measure of tranche riskiness such as ``moneyness'' or ``distance to default''.
Because of the ambiguity of such a procedure, the whole bespoke pricing methodology 
becomes quite {\it ad hoc}. In practice, it leads at times to negative bespoke 
tranche prices, which is not surprising given that interpolation in the base 
correlations space does not respect no-arbitrage constraints.

The approach based on the random thinning technique offers an alternative way
to bridge the gap between the index and bespoke portfolio. Consider a particular bespoke
that is obtained by substituting some name $ A $ in the index portfolio with 
another name $ B $ which is not a part of the index portfolio. The p-matrix 
corresponding to such a bespoke portfolio can be obtained in the same way as 
in the above calculation of single name sensitivities, i.e. we change a row $ i_0 $
corresponding to name $ A $ in such a way that the row is now calibrated to name
$ B $, and then rescale all columns in the p-matrix keeping the $ i_0$-th column
intact to retain the column constraints. The same idea can be applied in more 
complex situations as well, when the number of substituted names is larger 
than one, or when a name is added to the index portfolio rather than replacing
another name. We want to emphasize that this procedure, while being very simple, respects
no-arbitrage constraints and also makes sense in terms of adjustment to the risk
level of the bespoke portfolio. To illustrate this point, consider again our first example
with a substitution of one name. In this case, if the substituted name has a higher spread
than the withdrawn name, then $ (\delta p)_{in} > 0$ which translates into $ (\delta w)_n > 0 $,
i.e. tail probabilities increase. Therefore, our framework produces the correct directional
effect of the name substitution in the bespoke portfolio.
A more detailed analysis of implication of random thinning 
technique to the bespoke problem will be presented elsewhere.  

\subsection{Non-uniqueness of hedge ratios}

As should by clear by now from the previous sections, single name hedge 
ratios obtained within the top-down approach are not unique. Their 
non-uniqueness has two sources: the dependence on the initial guess in
the construction of TD-matrices, and non-uniqueness of the rule of tweaking 
these matrices.   
Non-uniqueness of hedges might appear to be a drawback 
of the top-down modeling paradigm. However, it is important to acknowledge that 
such a non-uniqueness is by no means specific to top-down models. 
In fact, it is exactly the same two sources of non-uniqueness of hedge ratios
that also exist in bottom-up models, once we move away from over-simplified 
static models such as the Gaussian copula model.
Indeed, consider e.g.
a dynamic bottom-up factor model where for a given name $ i $ the clean spread
$ \lambda_i $ is modeled as follows:
\beq
\label{bottomUp}
\lambda_i (t) = a_i \, \bar{\lambda}(t) + \varepsilon_i(t), 
\eeq
where $ \bar{\lambda}(t) $ is a common non-negative process, e.g. a 
Feller (CIR) diffusion,
and $ \varepsilon_i(t) $ is an idiosyncratic non-negative process (which can e.g. be 
another Feller diffusion). Let $ \vec{\bar{\Theta}} $ and $ \vec{\Theta}_i $ be 
vectors of parameters for these two processes, respectively. 
Calibration of the model amounts to choosing a set of 
parameters $ a_i $,  $ \vec{\bar{\Theta}}   $ and 
$ \vec{\Theta}_i  $ for all names $ i$ that provide the best fit to 
single names and tranches spreads. As a rule rather than exception, the resulting
objective function has multiple local minima, which in practice lead to
dependence on the initial guess for these parameters\footnote{In principle,
a global optimization algorithm would be able to find  the absolute minimum,
but such algorithms are rarely used in practice due to their lower speed.
Instead, local search algorithms are used, which generally 
retain the dependence on the initial guess.}.
As a result, calibration in a bottom-up model is typically non-unique
for all practical purposes.

The second source of non-uniqueness, namely, the non-uniqueness of a rule of tweaking
parameters, is also an issue for bottom-up models as well.
Indeed, for a given choice of parameters in (\ref{bottomUp}), 
there is an infinite number of possible ways a combination of tweaks of   
$ \vec{\bar{\Theta}}  $  and $ \vec{\Theta}_i  $ can be chosen such that they 
produce the same tweak of the CDS spread for name $ i $. 

Note that the choice of the 
split between tweaks of $ \vec{\bar{\Theta}}  $ and $ \vec{\Theta}_i  $ implies a 
particular way the correlation structure of the model changes upon a tweak of the CDS 
spread. Hence, while we calibrate the single name dynamics and the correlation 
structure  to single name  and tranche data, the resulting single name 
sensitivities are only unique up to specification of the law of {\it change} of 
the correlation structure.
This, of course, is not unexpected. Indeed, a single name delta is defined as 
a ratio of changes of a tranche MTM to a CDS MTM under a bump of the CDS spread, 
{\it provided nothing else changes}. But the latter notion is too vague: it can
mean e.g. that the absolute correlation level stays the same, or, probably 
more sensible,  that the tranche riskiness stays the same. In the 
latter case, we {\it have} to adjust the correlation
parameter when bumping a CDS spread as long as the base correlation curve is 
non-flat\footnote{The latter point is well known to practitioners, see e.g.
Ref.~\cite{McGinty} which discusses resulting ambiguities within the base correlation 
framework.}. Thus, in a general case, a bump of a CDS spread is accompanied by 
a rule of changing correlation parameters, i.e. correlation changes are driven
by spread changes.     

In practice, the correlation skew (in particular, ATM base correlation) is known to be 
negatively correlated with the index level, but not perfectly.
This implies two things. On the one hand, it means that the optimal way to define the rules 
for calculation of sensitivity parameters can (and should) be tuned using empirical 
correlation between the skew and index level, so that single name hedges will pick up a part 
of correlation risk attributable to (driven by) index level moves.  
On the other hand, this means that 
there is a residual correlation risk that cannot be hedged using single names alone.
To hedge this exposure, we should add some tranches to the hedge portfolio\footnote{For 
an analysis of correlation risk hedging in the context of a pure ``top'' model, see 
an interesting paper by Walker \cite{Walker}.}. If single name hedges are chosen optimally,
the notional amount of correlation hedges might be smaller than in a sub-optimal situation,
and hence the hedge will be cheaper. We hope to return to this problem elsewhere.    

\clearpage
\appendix
\def\thesection{A}	
\setcounter{equation}{0} 

\def\theequation{\thesection.\arabic{equation}}


\def\thesection{A}	
\setcounter{equation}{0}

\def\thesection{B}	
\setcounter{equation}{0}

\end{document}